\documentclass[]{aiaa-tc}

\usepackage{subfigure}
\usepackage{floatrow}
\usepackage{wrapfig}
\usepackage{amsfonts,amsmath,amsthm,mathrsfs}
\usepackage{amssymb}		
\usepackage{array}
\usepackage{multicol}
\usepackage{url}

\newcommand{\degrees}{\ensuremath{^\circ}}
\newfloatcommand{capbtabbox}{table}[][\FBwidth]

 \title{Anisotropic Boundary Layer Adaptivity of Multi-Element Wings}

 \author{
  Kedar C. Chitale\thanks{Postdoctoral Research Associate, Rensselaer Polytechnic Institute, AIAA Member, Tel.: +1-518-596-4750} \\ 
  {\normalsize\itshape
   MANE Dept., Rensselaer Polytechnic Institute, Troy, NY 12180-3590}\\
   \and
     Michel Rasquin\thanks{Postdoctoral Research Associate, Argonne National Laboratory} \\
      {\normalsize\itshape Leadership Computing Facility, Argonne National Laboratory, IL 60439}
  \and
  Onkar Sahni\thanks{Assistant Professor, Rensselaer Polytechnic Institute, AIAA Senior Member}
\ and Mark S. Shephard\thanks{Professor, Rensselaer Polytechnic Institute, Tel.: +1-518-276-6795 }\\
  {\normalsize\itshape
  MANE Dept., Rensselaer Polytechnic Institute, Troy, NY 12180-3590} 
  \and
   Kenneth E. Jansen\thanks{Professor, University of Colorado Boulder, AIAA Associate Fellow, Tel.: +1-303-492-4359}\\
   {\normalsize\itshape
   Dept. of Aerospace Engineering Sciences, University of Colorado Boulder, Boulder, CO 80309-0429,  }\\
 }

 \AIAApapernumber{YEAR-NUMBER}
 \AIAAconference{Conference Name, Date, and Location}
 \AIAAcopyright{\AIAAcopyrightD{YEAR}}

\begin{document}

\maketitle

\begin{abstract}

Multi-element wings are popular in the aerospace community due to their high lift performance. Turbulent flow simulations of these configurations require very fine mesh spacings especially near the walls, thereby making use of a boundary layer mesh necessary. However, it is difficult to accurately determine the required mesh resolution {\it a priori} to the simulations. In this paper we use an anisotropic adaptive meshing approach including adaptive control of elements in the boundary layers and study its effectiveness for two multi-element wing configurations. The results are compared with experimental data as well as nested refinements to show the efficiency of adaptivity driven by error indicators, where superior resolution in wakes and near the tip region through adaptivity are highlighted. 

\end{abstract}

\section*{Nomenclature}

\begin{tabbing}
  XXXXX \= \kill
  $ y^+$ \> Dimensionless distance from the wall ($ u_\tau y/\nu$) \\
  $ Re$ \> Reynolds number \\
  $ C_p$ \> Coefficient of pressure \\
  $ C_L$ \> Coefficient of lift \\
  $ C_D$ \> Coefficient of drag \\
  $ b$ \> Span of the wing ($m$) \\
  $ \alpha$ \> Angle of attack (AoA) \\
   $ c_{local} $ \> Local chord ($m$)

 \end{tabbing}

\section{Introduction}

The accuracy of CFD simulations strongly depends on the mesh resolution and quality. In complex flow problems, it is difficult to determine the adequate mesh resolution {\it a priori}. In such cases, an initial mesh is used to get an estimate of the flow solution, and the mesh is adapted using {\it a posteriori} error indicators. In order to expedite the numerical computation, the resolution needs to be applied in a local fashion, which can be achieved by locally modifying the mesh elements, based on a size field. Traditionally, scalar error indicators are used which leads no change in mesh anisotropy. But many complex flow problems exhibit highly anisotropic features such as boundary layers in viscous flows and shock waves in high-speed flows. These features are most efficiently resolved with anisotropic elements, i.e., with elements that are stretched and oriented in a certain manner. 

For wall-bounded turbulent flows, boundary layers need to be resolved efficiently and accurately as they are the most prominent flow feature. Additionally when the boundary layers are turbulent, which is often the case for high Reynolds number flows, mesh spacing needs to be properly controlled. Meshing boundary layer region with isotropic elements will put an excessive demand on the computational resources due to an extremely large mesh. Furthermore, a fully unstructured anisotropic mesh results in poorly shaped elements (e.g., elements with aspect ratio above 1000) and in-turn leads to a numerical solution of poor quality\cite{Sahni}. To remedy these problems, layered, orthogonal and graded elements are used near the walls whereas the rest of the domain is filled with unstructured elements; this is referred to as a {\it boundary layer mesh}. A common method to construct such boundary layer meshes is {\it advancing layers method}\cite{Pirzadeh, Garimella, ItoAIAA}. Such a mesh provides a useful way to achieve very fine mesh spacings normal to the walls that are required for accurate turbulence modeling. As one moves away from the wall, the wall normal mesh spacings can be appropriately coarsened in a progressive manner. Due to these properties, most of the unstructured grid turbulent flow simulations use some form of such specially constructed boundary layer meshes. 

During adaptivity, it is desirable to maintain this layered structure of elements inside the boundary layer. Adaptation procedures based on local mesh modifications for boundary layer meshes have been presented in previous work\cite{Sahni}. These procedures have recently been extended to work in parallel for distributed memory computer systems on thousands of cores\cite{Ovcharenko}. For turbulent flows, proper care needs to be taken while meshing the boundary layers with regards to the thickness specification. Different classes of turbulence models have different requirements on the wall-normal spacing. Boundary layer adaptivity has the challenging task of either preserving or adapting the thickness structure without violating the conditions dictated by the turbulence models. Boundary layer adaptivity in the thickness direction using physics-based indicators has been carried out for simpler configurations\cite{Chitale}.

Anisotropic boundary layer mesh adaptivity is significantly more complex for configurations, such as multi-element wings, than previously studied examples. The challenges are two fold: the surface mesh needs to be adapted anisotropically which presents challenges for complex geometries, and the geometric approximation of the mesh needs to be improved as the mesh is refined which is critical for curved surfaces. Adaptive analysis of multi-element wings has previously been carried out\cite{ParkHiLift} but some part of the boundary layer mesh had been frozen to get around some of the challenges presented by surface adaptation. In this paper we explore boundary layer adaptivity of multi-element wings along with surface adaptation and improving the geometric approximation. Here, we focus on in-plane adaptation of the boundary layer meshes.

The article is organized as follows. Section~\ref{s:AnisoAdapt} describes anisotropic adaptivity techniques and how they are extended to boundary layers. Section~\ref{s:results} illustrates the capability of our approach by showing results for two cases both involving multi-element wings. Comparisons are made with experimental data as well as nested refinement approach to show the efficiency of error-indicator driven adaptivity.



\section{Anisotropic adaptivity and boundary layers}
\label{s:AnisoAdapt}

Boundary layer meshes are widely used in simulations of turbulent flows. These semi-structured meshes provide an easy way to achieve elements with anisotropy of 10,000 or even greater, without creating badly shaped elements. This helps in accurately resolving the boundary layers, which are often turbulent in nature for high Reynolds number flows. 

\begin{figure}[h!]
\begin{center}
\subfigure[Schematic of the boundary layer mesh] {
 \includegraphics[width = 7 cm]{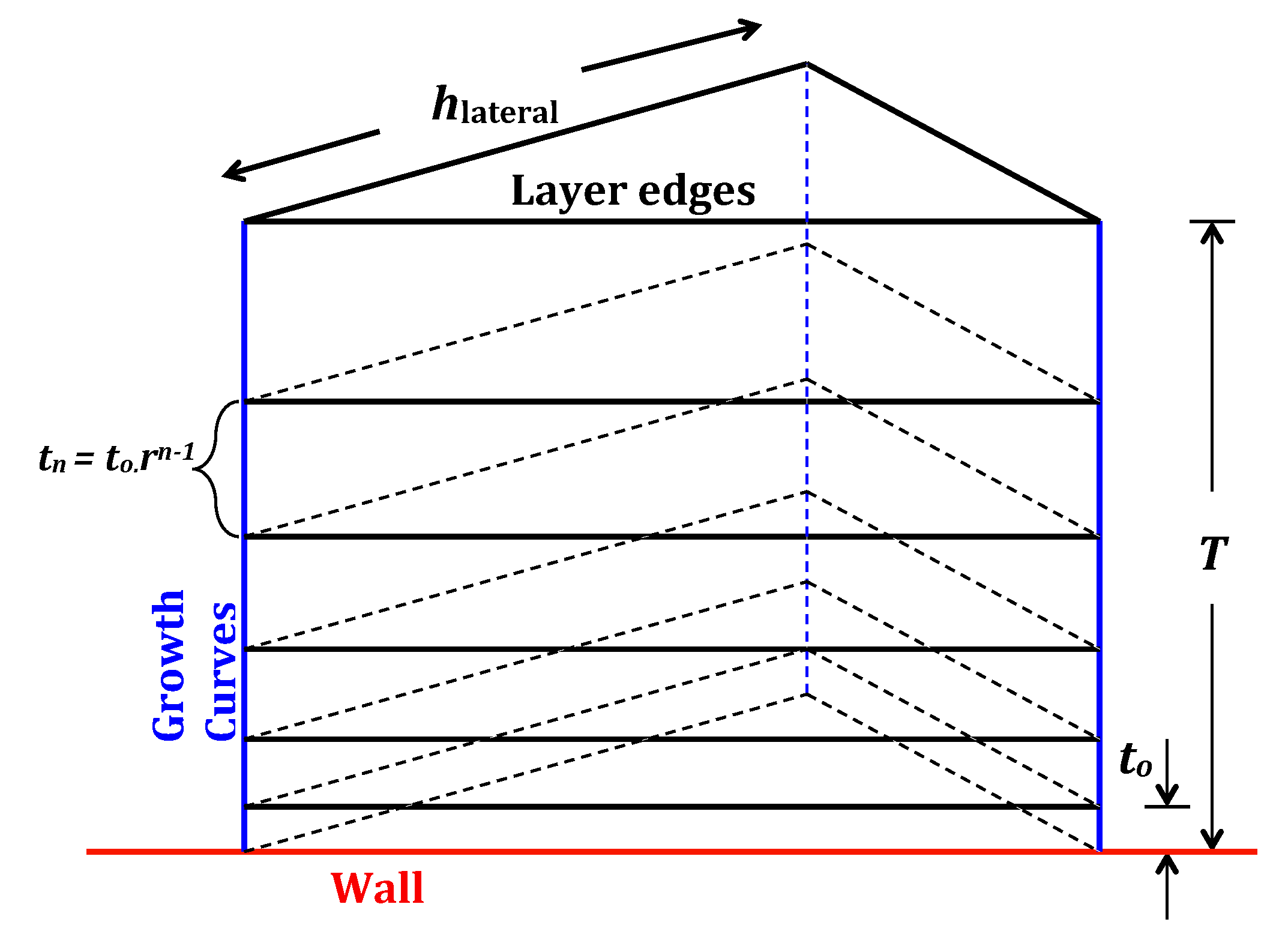}
 \label{f:BLStructure}
}
\subfigure[Boundary layer mesh for a pipe geometry] {
 \includegraphics[width = 8 cm]{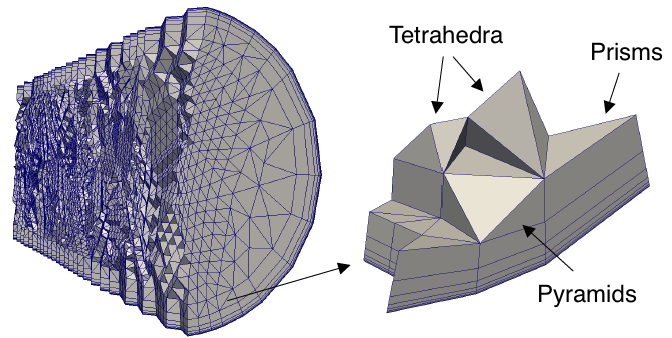}
 \label{f:PipeBL}
 }
 \vspace{-10pt}
 \caption{Boundary layer mesh}
 \label{f:MeshBL}
 \end{center}
\end{figure}

Figure~\ref{f:PipeBL} shows an example of a boundary layer mesh for a simple pipe geometry. For 3-D meshes, boundary layer meshes are comprised of prisms and pyramids (considering triangular surface mesh), whereas the unstructured region is meshed with tetrahedra. 

\subsection {Hessian driven adaptivity}

Outside the boundary layer, the mesh can also be anisotropic. Since the level of anisotropy required outside the boundary layer is much less, general unstructured anisotropic meshes are used and the mesh anisotropy is defined using the well-known Hessian (or interpolation error) based methods\cite{Buscaglia, Diaz, PerPei92, Pain01}. The anisotropic adaptive meshing procedures used in this work, is based on the local modifications of the mesh elements following a {\it mesh metric field}\cite{XLi2}. The mesh metric is derived from a Hessian matrix, which is a symmetric matrix constructed from the second derivatives of the flow solution variables. Traditionally speed and density are chosen as the solution variables, but a combination of different variables can also be used. It is possible to obtain local estimates of the interpolation error in different norms, based on the Hessian matrix\cite{Apel}. 

The Hessian matrix is decomposed as $ H = R\Lambda R^T$, where $ R$ is the matrix of eigenvectors and $ \Lambda$ is the diagonal matrix of eigenvalues. The directions associated with the eigenvectors are referred to as the principal directions and the eigenvalues are equivalent to the second derivatives along these directions. High eigenvalues are associated with high error in the corresponding principal direction. Similarly, a low eigenvalue means lower error in the corresponding direction \cite{Sahni2}. Mesh sizes (mesh edge lengths in particular directions) can be calculated as scaled inverses of the eigenvalues at each vertex of the mesh.

\begin{figure}[h!]
\begin{center}
\vspace{10pt}
\includegraphics[width=9 cm]{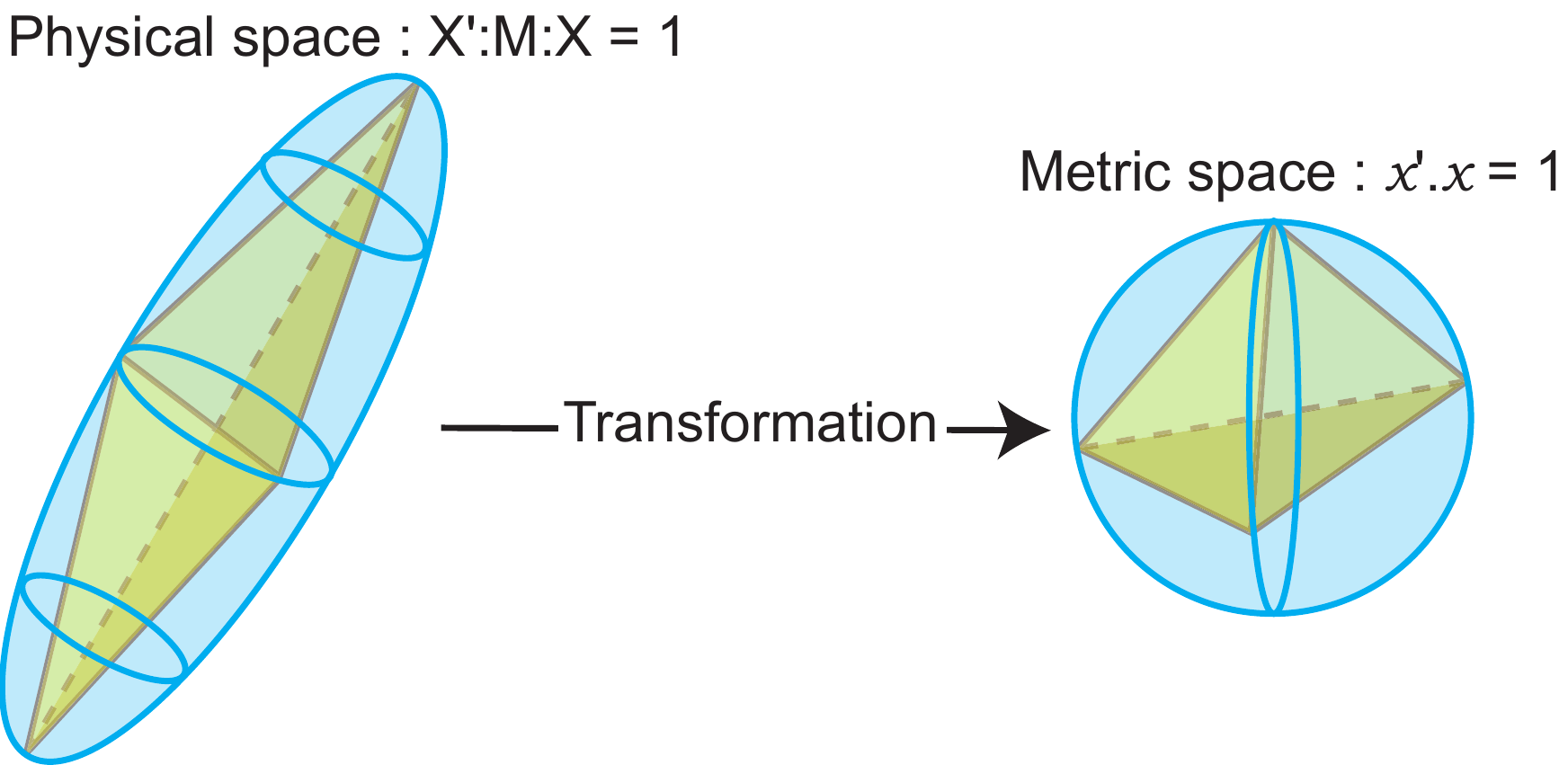}
\caption{Transformation associated with the mesh metric tensor\cite{Sahni}} 
\label{f:MeshMetricTransform}
\end{center}
\vspace{-10pt}
\end{figure}

The mesh metric field can be thought of as a transformation matrix which defines a mapping of an ellipsoid in the physical space into a unit sphere in the metric space, as shown in Figure~\ref{f:MeshMetricTransform}. An element of any shape in the physical space is transformed to an equilateral element in the metric space with this transformation. The goal of the mesh adaptation algorithm is to achieve unit edge lengths in the metric space. For meshes of complex domains, this criteria is usually relaxed to constrain edge lengths in the metric space to be within an interval around $1$ (i.e., unit length)\cite{Diaz,XLi2}.

\subsubsection{Extension to boundary layers}

The methodology outlined above works well for unstructured elements. When working with boundary layers we want to preserve their layered nature, and using this technique directly does not guarantee that. To extend anisotropic adaptivity to boundary layer meshes we instead use the approach described below.

\begin{figure}[h!]
\begin{center}
\vspace{10pt}
\subfigure[Decomposition of the boundary layer mesh] {
\includegraphics[width=7.5cm]{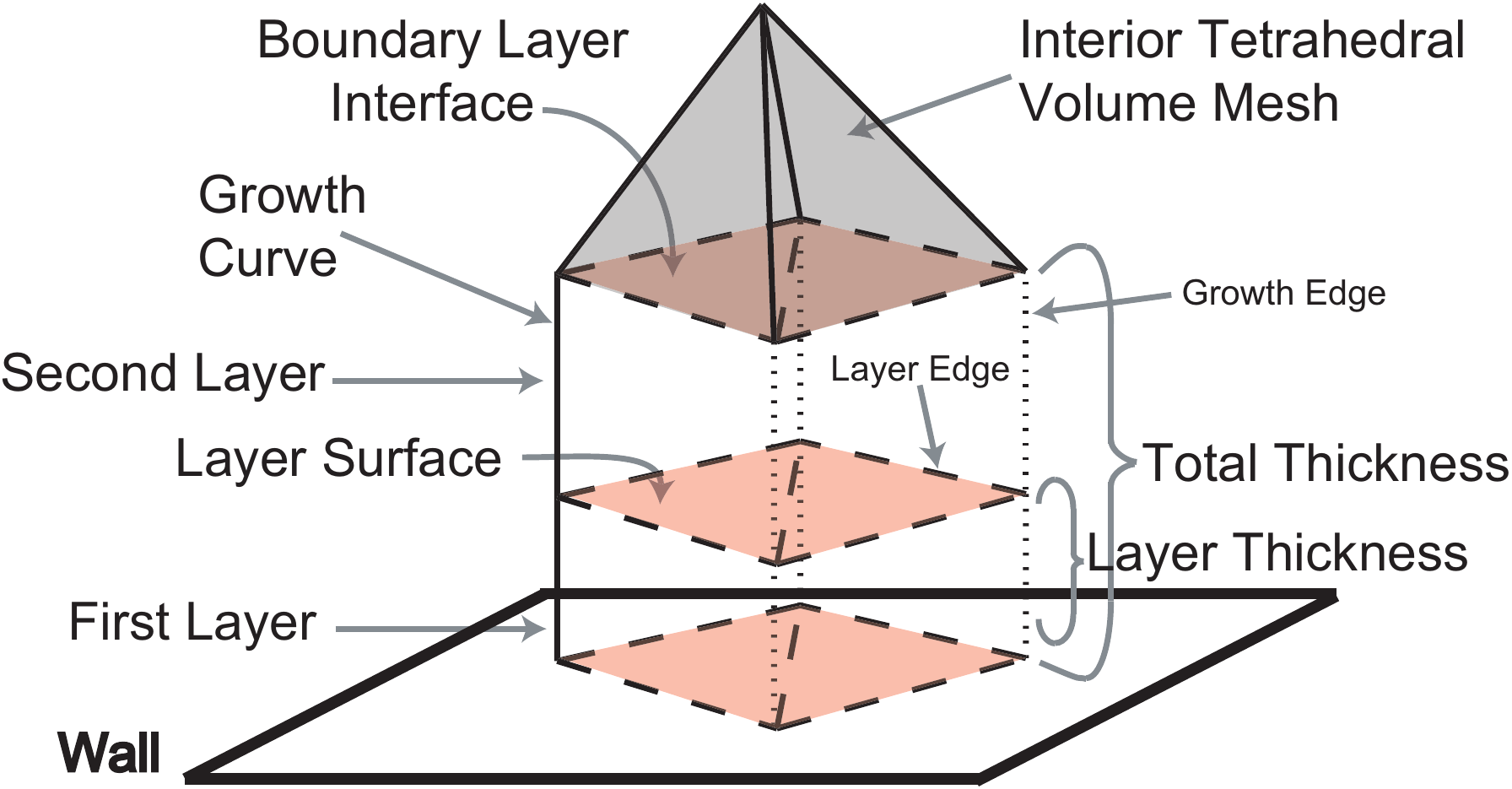}
\label{f:BLDecompose}
}
\subfigure[Decomposition of the ellipsoid]{
\includegraphics[width=8cm]{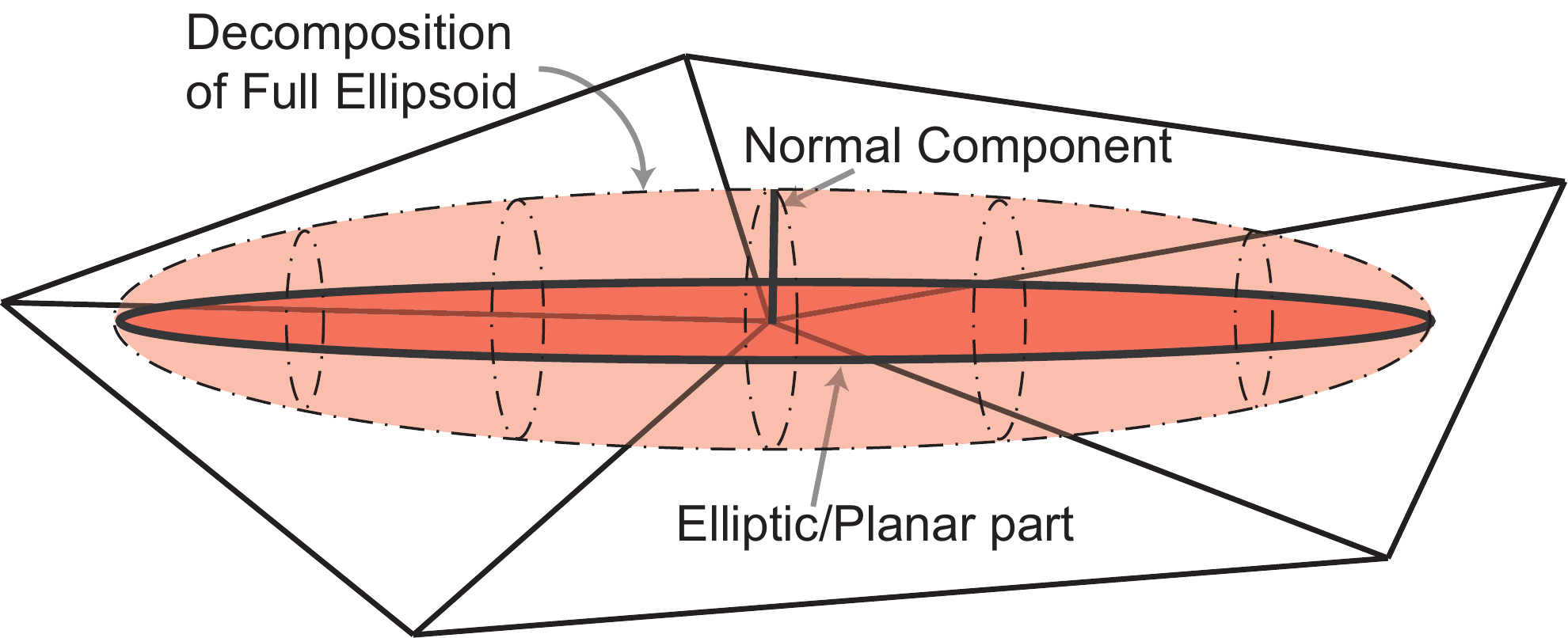}
\label{f:EllipsoidDecompose}
}
\vspace{-10pt}
 \caption{Conceptual extension of the mesh metric to a boundary layer mesh\cite{Sahni}}
 \label{f:MeshMetricAndEquation}
\end{center}
\end{figure}

Figure~\ref{f:BLDecompose} shows a conceptual decomposition of the boundary layer mesh. The boundary layer meshes can be viewed as a product of a layer surface (2D) and a thickness (1D) mesh. The lines which are orthogonal to the wall are referred to as the {\it growth curves}, and the triangular surfaces parallel to the wall are referred to as the {\it layer surfaces}. Each layer of elements is formed with the help of the layer surfaces above and below, connected by the growth edges in between. The mesh size on the layer surfaces is referred to as the {\it in-plane} or lateral size and that on the growth curves is referred to as the {\it normal} spacing or thickness. The ellipsoid in Figure~\ref{f:MeshMetricTransform} can be decomposed as an ellipse projected on the layer surface and a normal component aligned with the growth curve. This concept is shown in Figure~\ref{f:EllipsoidDecompose}. 

Adaptivity is carried out in two stages: {\it in-plane adaptation} that achieves the required mesh sizes on the layer surfaces and does not affect the thickness, and {\it thickness adaptation} which changes the normal spacing of the boundary layers. The in-plane adaptation is driven by the mesh metric field calculated from the Hessian as described in this section (see Sahni et al.\cite{Sahni2} for more details). 

\subsection{Combined strategy for adaptivity}
\label{ss:combined}

Though the Hessian is very useful in extracting the directional information from the flow solution, it is not always accurate. For example, an extrapolation technique is used near the walls to project the interior values of the gradients on to the nodes that lie on the domain boundaries. For high Reynolds number viscous flows velocity gradients near the wall region are very high. When using Hessian of speed variable, it is seen that the near wall resolution requests are often very small and can introduce excessive refinement.
 
To remedy these problems, some {\it a posteriori} error estimators can be used in combination with the Hessian to drive the adaptation. Vendetti et al.\cite{Vendetti2} developed an approach, wherein Hessian of Mach number sets the anisotropy (i.e., aspect ratio) of the mesh elements and the adjoint adaptation parameter dictates the smallest mesh spacing in the local mesh metric tensor. For $ C^0$ finite elements, calculating the adjoint adaptation parameter requires solving another system of partial differential equations (PDEs) to complete the dual problem. This can be expensive in terms of the computational resources.

Another option is to use explicit error indicators based on residuals of the PDEs (i.e., strong form residuals), which can be computed easily. Furthermore, root-mean-square of the solution fluctuations can also be used and readily obtained from the solver. These quantities have been traditionally used as scalar error indicators to drive isotropic adaptivity, which results in equilateral elements. In this study, we have implemented an approach such that the Hessians provide the relative scales and the directional information for adaptation, but the local smallest mesh size (at any point) is calculated from the error indicators based on PDE residual or root-mean-square of fluctuations. These error indicators are assigned relative weights and are combined to give the final scalar indicator, which drives the local smallest mesh size during adaptivity. 

The above approach has a distinct advantage in aerodynamic flows. Usually Hessians of speed or pressure are not very dominant in the wake. With velocity Hessians, most of the adaptivity is concentrated on the walls and with pressure Hessians, stagnation region experiences a lot of mesh refinement. Due to this, the wake region usually remains under resolved when (only) Hessians are used for adaptation. However, PDE residuals are usually significant in the wake region and can be a better indicator of the error in this region. Using a combination of the Hessians and scalar error indicators ensures that anisotropy is not lost and yet important flow features, such as wakes and tip vortices, receive needed refinement. 


\section{Results}
\label{s:results}

The flow solver used in this work is based on the SUPG finite element formulation of the incompressible Navier-Stokes equations\cite{WhiJan99}. $C^0$ piecewise linear finite elements are used for analysis. The time integration is based on an implicit generalized-$\alpha$ method\cite{JanWhi99}. The solver supports various turbulence models like RANS, LES, DES. The flow simulations in this work were performed with steady-state RANS-SA (Spalart-Allmaras)\cite{SA} turbulence model.

In this section we present results for two complex cases both involving multi-element wings, where the above mentioned adaptive boundary layer meshing techniques are employed. The first case is NASA trap wing which was the focus of $1^{\text{st}}$ high-lift prediction workshop\cite{HLPW1}. The second case is the EUROLIFT DLR-F11 configuration which was the focus of recently conducted $2^{\text{nd}}$ high-lift prediction workshop\cite{HLPW2}. 

\begin{figure}[h!]
\begin{center}
\vspace{10pt}
\subfigure[NASA trap wing]{
 \includegraphics[width = 8.6cm]{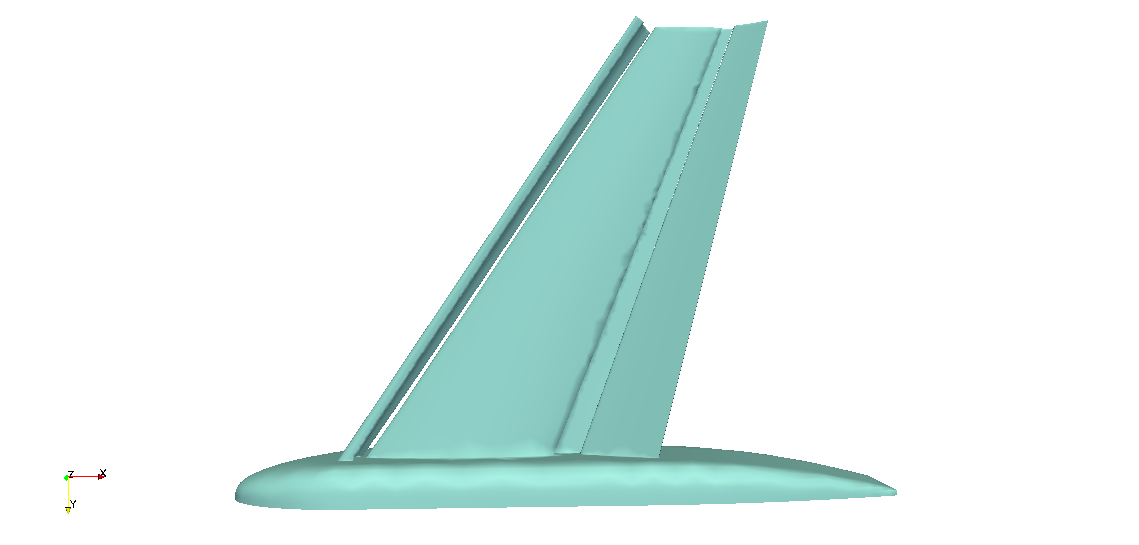}
 \label{f:TrapwingGeom}
 }
 \hspace{-50pt}
 \subfigure[EUROLIFT DLR-F11 high lift configuration]{
 \includegraphics[width=8.6cm]{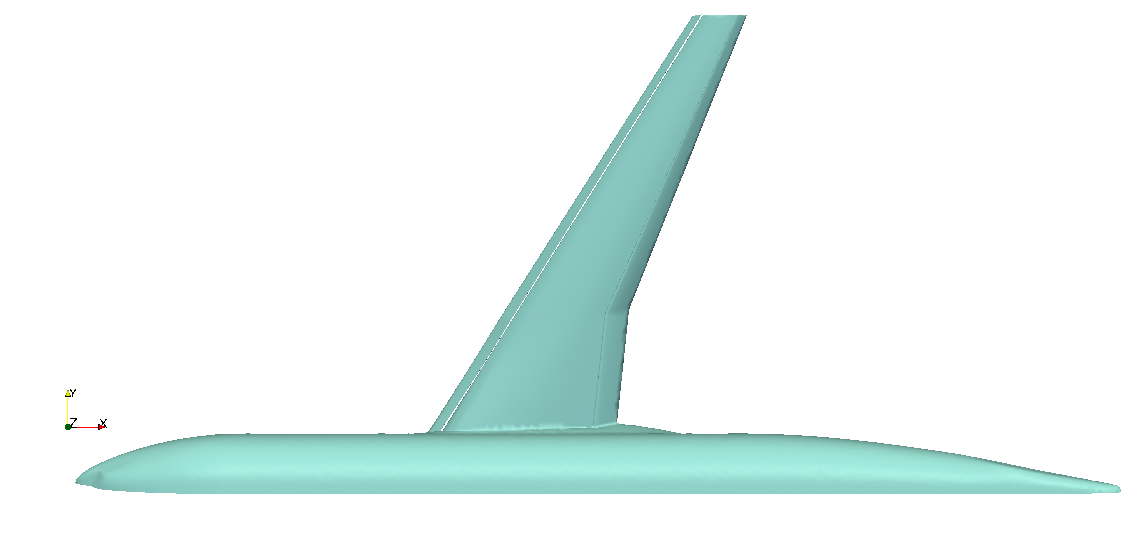}
  \label{f:DLRGeom}
 }
  \vspace{-10pt}
 \caption{Geometries of the two multi-element wings}
 \label{f:TWGeom}
\end{center}
\end{figure}

The various meshes in the analysis are labeled with `LEV', indicating the level of the adapted mesh, with LEV0 as the initial mesh through LEV3 which is the third adapted mesh. The adapted mesh results are compared with nested refinement mesh results. These meshes are labeled with `NLEV', indicating the nested refinement level. 
 
 \subsection{NASA trap wing}

The first case for boundary layer mesh adaptation is a trapezoidal wing geometry. This multi-element wing has slats and flaps in addition to the main wing, and was used as the benchmark case in the $1^{\text {st}}$ high-lift prediction workshop organized by NASA in 2010~\cite{HLPW1}. The configuration used is a landing configuration with the slat at $30\degrees$  and the flap at an angle of $25\degrees$  with respect to the chord of the main element. The experiments for NASA trap wing were performed at NASA Langley in a $14\times22$-foot wind tunnel \cite{TrapWingExp}.

For analysis, an initial mesh was generated with mixed element boundary layers and used to obtain an initial solution. Two adaptivity loops were carried out and a combined approach of Hessians and scalar error indicators was used to drive adaptivity as explained in Section~\ref{s:AnisoAdapt}.\ref{ss:combined}. 

The flow was modeled as an incompressible turbulent flow with wall-resolved Spalart Allmaras turbulence model \cite{SA}. Other relevant case setup information is given in Table~\ref{t:TWSetup}.

\begin{table}[h!]
\centering
\newcolumntype{A}{>{\centering\arraybackslash}m{3 cm}}
\newcolumntype{B}{>{\centering\arraybackslash}m{4 cm}}
\newcolumntype{D}{>{\centering\arraybackslash}m{2 cm}}

      \begin{tabular}{|A|B|A|A|}
      \hline
  	 Mach number & Mean aerodynamic chord (MAC) & $\mathrm{Re_{MAC}}$ & Angle of attack  \\ \hline
	0.2 &  1.0 $ m$ & 4.3 million &13\degrees  \\ \hline
      \end{tabular}
  \caption{Problem definition for NASA trap wing}
  \label{t:TWSetup}
\end{table}

\begin{figure}[h!]
\begin{center}
\vspace{10pt}
 \includegraphics[width = 16 cm]{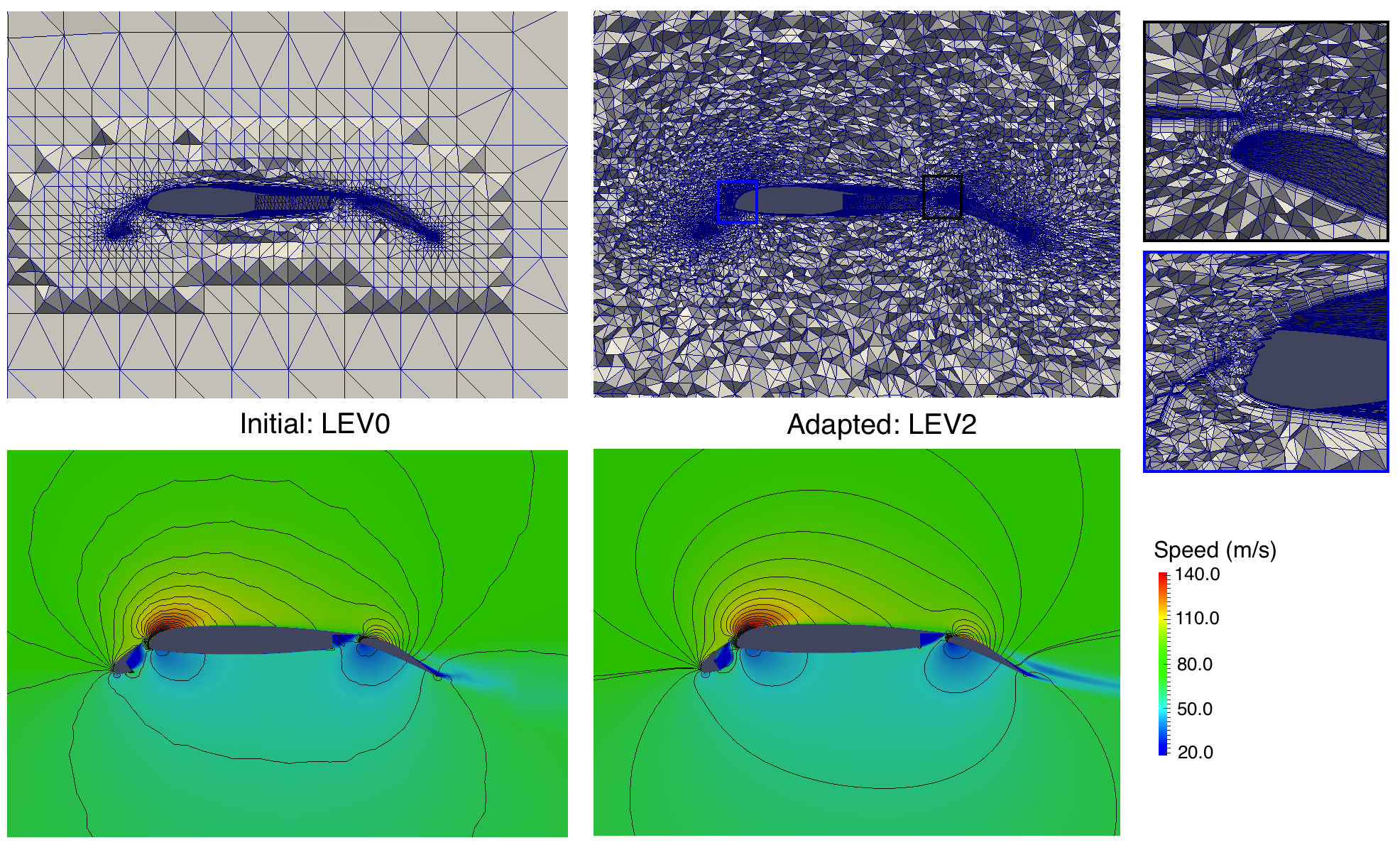}
 \caption{Cuts of the initial LEV0 and adapted LEV2 meshes, showing anisotropic adaptivity and speed distribution along with pressure contours}
 \label{f:TWMeshCut}
\end{center}
\end{figure}


The initial mesh for NASA trap wing has 3.3 million elements, the adapted LEV1 mesh has 7.8 million elements and the adapted LEV2 mesh has 12.3 million elements. The seemingly small increase in the number of elements per adapt cycle can be attributed to the anisotropy in the elements. Figure~\ref{f:TWMeshCut} shows cuts of the initial LEV0 and the adapted LEV2 meshes and corresponding speed distribution (colored) and pressure contours (black lines). Superior capturing of the wake with adaptivity is evident from the anisotropic mesh in this region and the speed pictures. The pressure contours which are a bit jagged for the initial mesh, become more regular and smooth in the adapted mesh. The zooms on the right show the adapted mesh near the leading edges of the main wing and the flap element showing the refinement in these regions, also displaying refinement near the trailing edges of the slat and the main wing.

\begin{figure}[h!]
\begin{center}
\subfigure[Initial mesh: LEV0] {
	\includegraphics[width=7.2cm]{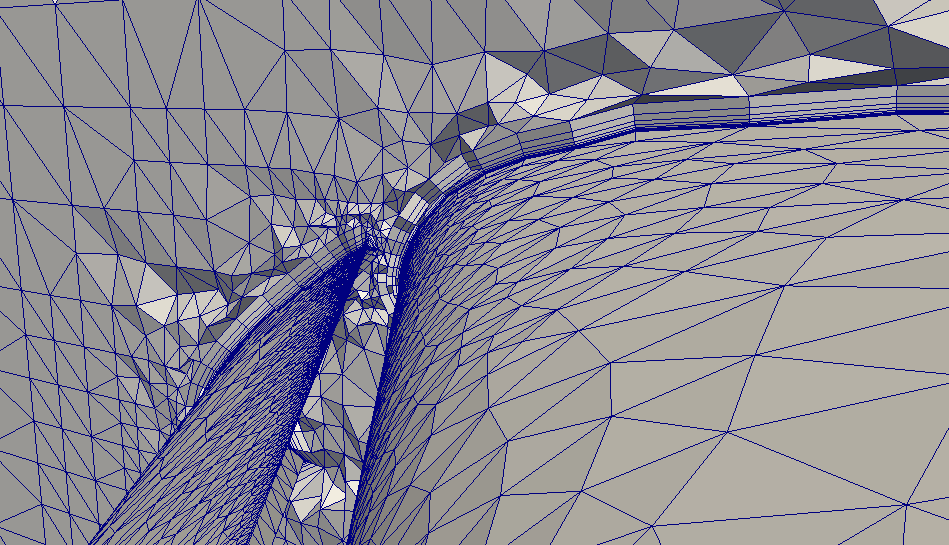}
	\label{f:TWLeadingInit}
}
\subfigure[Adapted mesh: LEV2] {
	\includegraphics[width=7.2cm]{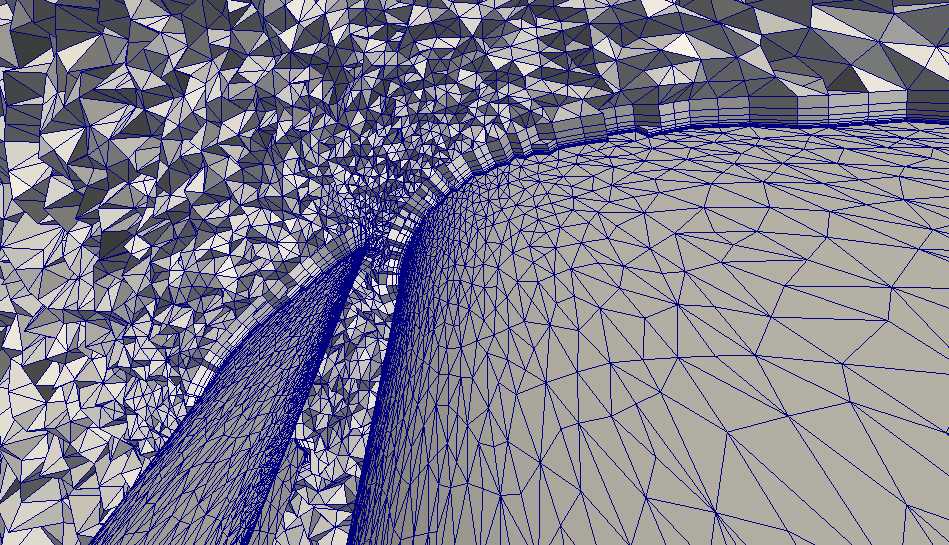}
	\label{f:TWLeadingAdapt2}
}
 \vspace{-15pt}
 \caption{Cut view of the initial (LEV0) and adapted (LEV2) meshes near the leading edge of the main element of NASA trap wing}
 \label{f:TELeadingEdge}
\end{center}
\end{figure}

Figure~\ref{f:TELeadingEdge} displays the mesh near the leading edge of the main wing for the initial LEV0 and the adapted LEV2 meshes to illustrate the anisotropy developed in the spanwise direction. The leading edge is refined such that the elements have smaller edge lengths in the streamwise direction as compared to the spanwise direction. This behavior is expected because the flow changes in the streamwise direction are very large near the leading edge of the main wing. This figure also shows the in-plane boundary layer adaptivity as the stacks of boundary layers get split in the layer direction, especially near the nose. 

\begin{figure}[h!]
\begin{center}
\includegraphics[width=14cm] {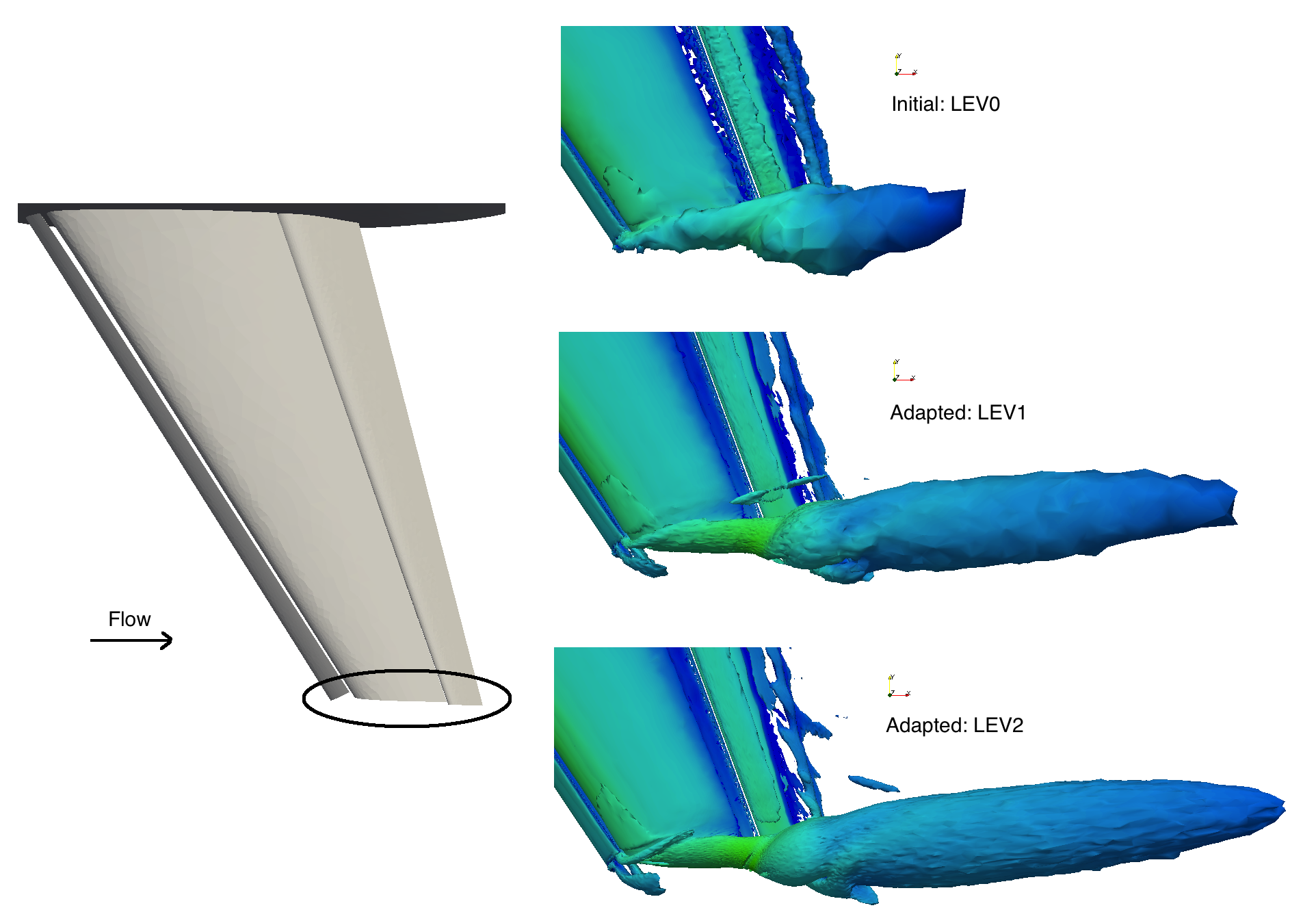}
\vspace{-10pt}
 \caption{Iso-surfaces of the Q criterion colored by speed for NASA trap wing showing the tip vortex}
 \label{f:TWQIso}
\end{center}
\end{figure}

Figure~\ref{f:TWQIso} shows the iso-surfaces of the Q criterion which capture the elongated tip vortex. The Q criterion is calculated as: $ 0.5(\Omega_{ij} \Omega_{ij} - S_{ij} S_{ij})$, where $\Omega$ is the asymmetric part and $ S $ is the symmetric part of the velocity gradient tensor. Since vorticity usually captures high normal gradients in boundary layer flows, the vortical structures get overshadowed. The Q criterion eliminates this problem by signifying the regions where rate of rotation dominates the shear strain rates, which is often the case for the tip vortex for positive Q values.

The iso-surfaces indicate that the tip vortex dissipates quickly for the initial mesh due to its coarseness. However, its real structure is captured much better in the adapted meshes. This is important because it confirms that  the tip area and the wake are resolved because of our error indicator driven adaptivity and so the flow structures are captured much better in this region than the coarse initial mesh. The effect of anisotropic elements in the wake can also be seen on the structure of the tip vortex.

\begin{figure}[h!]
\begin{center}
\subfigure{
	\includegraphics[width=5.2cm]{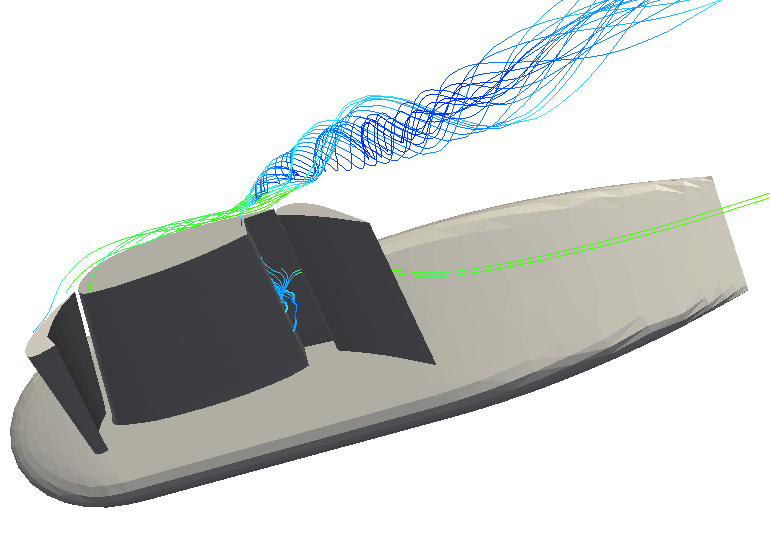}
	\label{f:TWTipStreamInit}
}
\subfigure{
	\includegraphics[width=5.2cm]{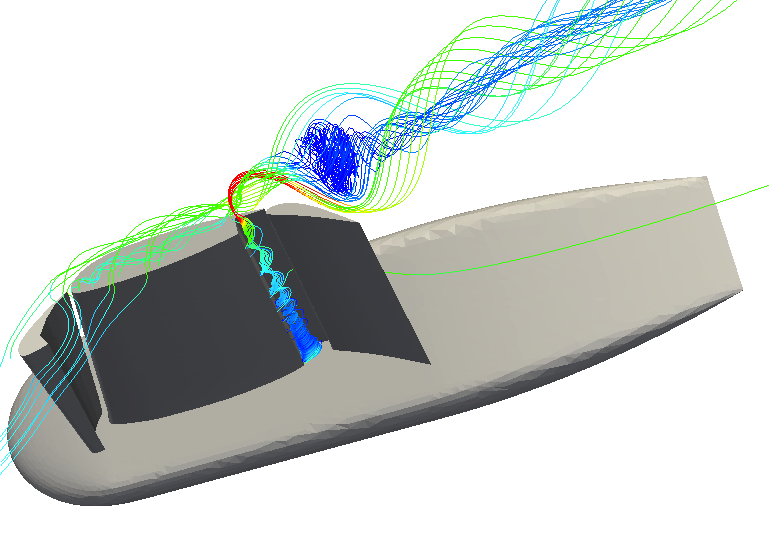}
	\label{f:TWTipStreamAdapt1}
}
\subfigure{
	\includegraphics[width=5.2cm]{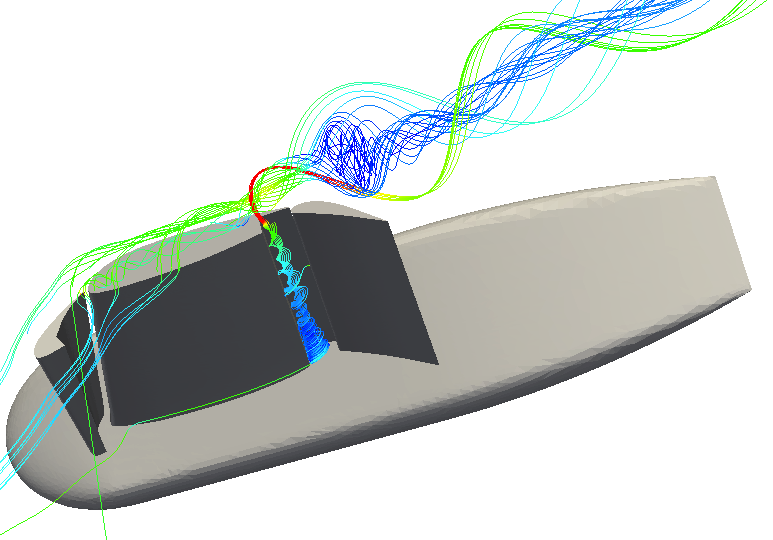}
	\label{f:TWTipStreamAdapt2}
}

\subfigure[Initial mesh: LEV0] {
	\includegraphics[width=5cm]{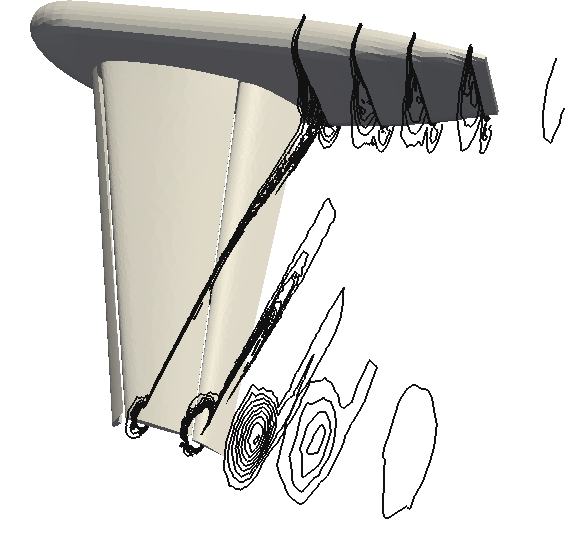}
	\label{f:TWVortContInit}
}
\subfigure[Adapted mesh: LEV1] {
	\includegraphics[width=5cm]{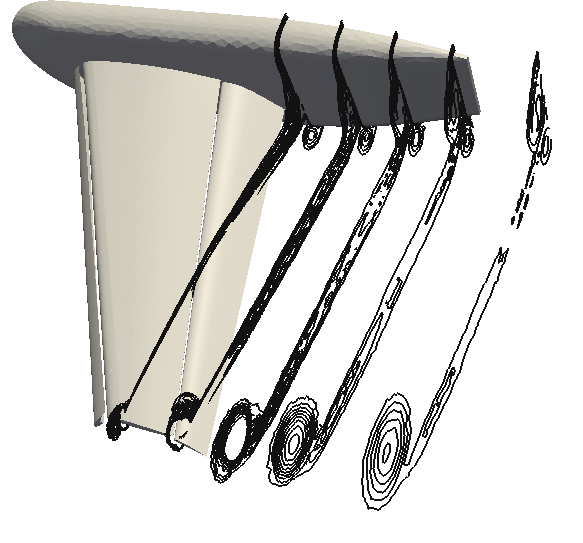}
	\label{f:TWVortContAdapt1}
}
\subfigure[Adapted mesh: LEV2]{
	\includegraphics[width=5cm]{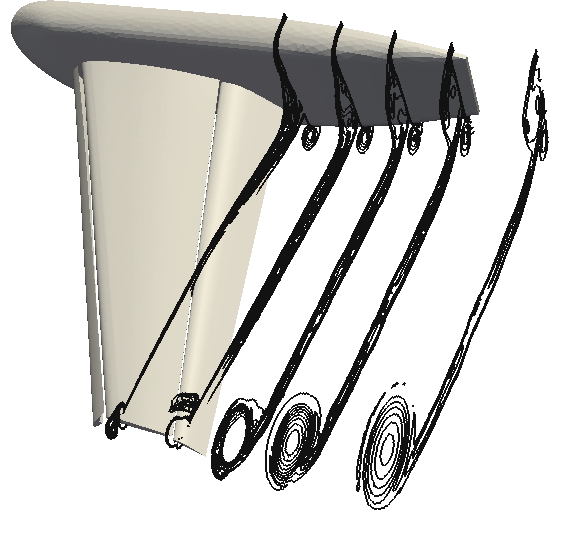}
	\label{f:TWVortContAdapt2}
}
\vspace{-10pt}
 \caption{On top: streamlines near the tip, on bottom: contours of vorticity on various slices with normals in the flow direction for NASA trap wing}
 \label{f:TWStreamAndVort}
\end{center}
\end{figure}

Results over a larger portion of the flow domain are shown in Figure~\ref{f:TWStreamAndVort} which displays the streamlines near the tip and vorticity contours on the bottom for different meshes. The initial coarse mesh gives a lot of separation and a dominant free shear layer near the tip, but the adapted meshes predict more turbulent activity in this region. The streamlines inside the flap slot of the main wing are not clearly seen in the initial mesh but can be seen in the adapted meshes. The vorticity contours of the initial mesh indicate that the tip vortex is not captured properly due to the coarse nature of the initial mesh (LEV0). The initial mesh is not able to completely resolve the side-of-body vortices either, which are resolved much better in both of the adapted meshes. Overall, adaptivity captures the tip and side-of-body vortical features much better due to increased refinement in appropriate regions. 

\begin{figure}[h!]
\begin{center}
\hspace{-25pt}
\subfigure[Slat element: 17\% span] {
	\includegraphics[width=5.4cm]{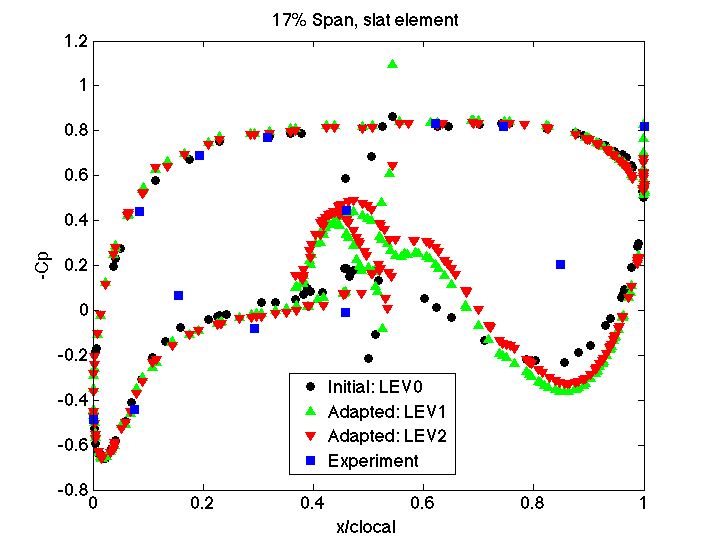}
	\label{f:TWSlat17}
}
\hspace{-25pt}
\subfigure[Slat element: 50\% span] {
	\includegraphics[width=5.7cm]{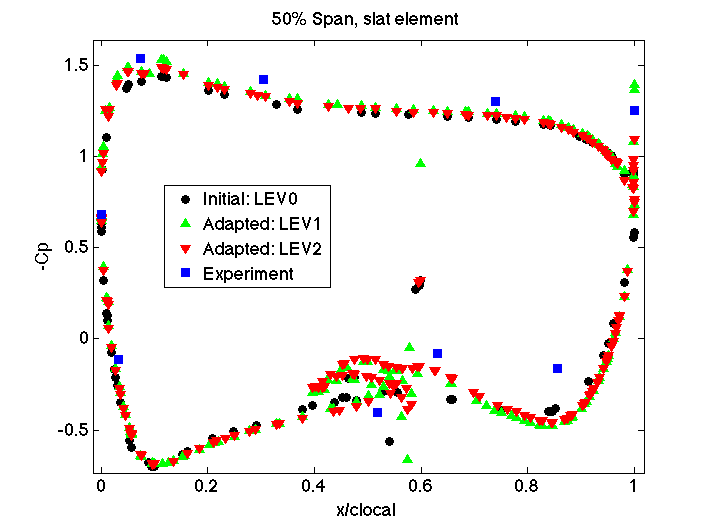}
	\label{f:TWSlat50}
}
\hspace{-25pt}
\subfigure[Slat element: 98\% span]{
	\includegraphics[width=5.7cm]{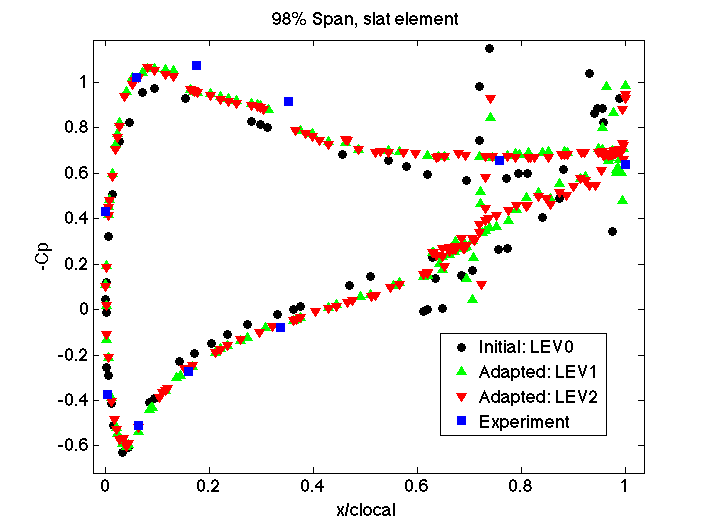}
	\label{f:TWSlat98}
}
\hspace{-25pt}

\subfigure[Main element: 17\% span] {
	\includegraphics[width=5.6cm]{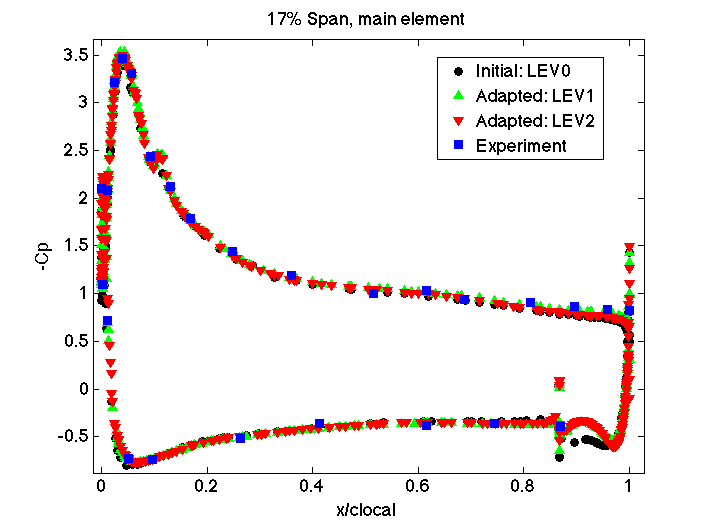}
	\label{f:TWMain17}
}
\hspace{-25pt}
\subfigure[Main element: 50\% span]{
	\includegraphics[width=5.6cm]{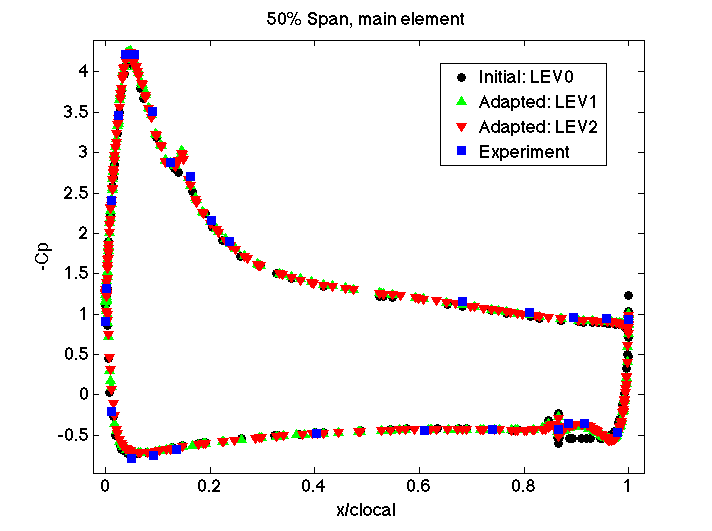}
	\label{f:TWMain50}
}
\hspace{-25pt}
\subfigure[Main element: 98\% span]{
	\includegraphics[width=5.6cm]{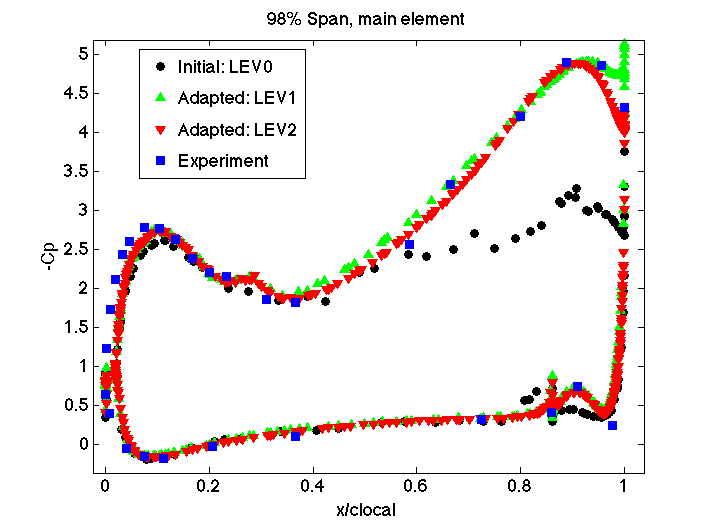}
	\label{f:TWMain98}
}	

\subfigure[Flap element: 17\% span] {
	\includegraphics[width=5.6cm]{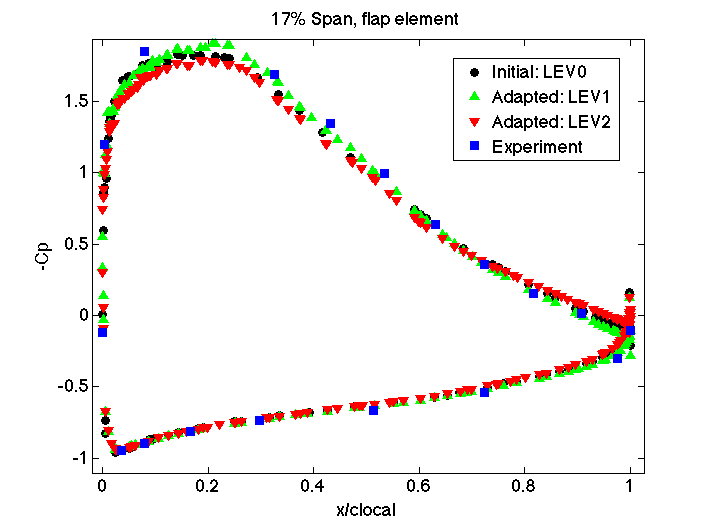}
	\label{f:TWFlap17}
}
\hspace{-25pt}
\subfigure[Flap element: 50\% span]{
	\includegraphics[width=5.6cm]{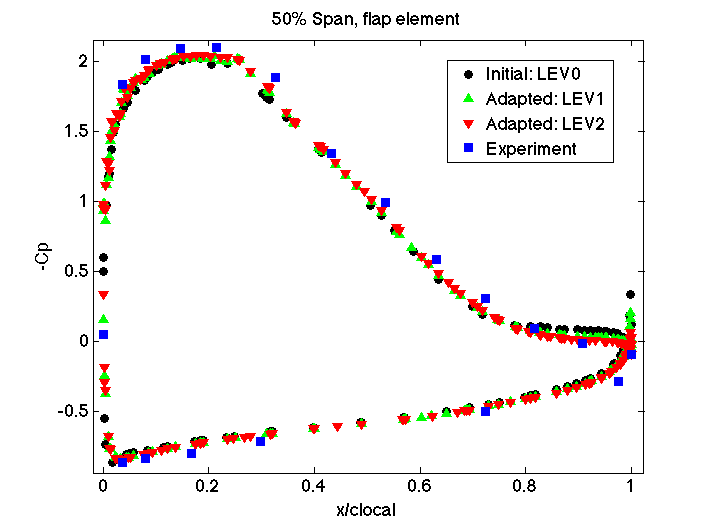}
	\label{f:TWFlap50}
}
\hspace{-25pt}
\subfigure[Flap element: 98\% span]{
	\includegraphics[width=5.6cm]{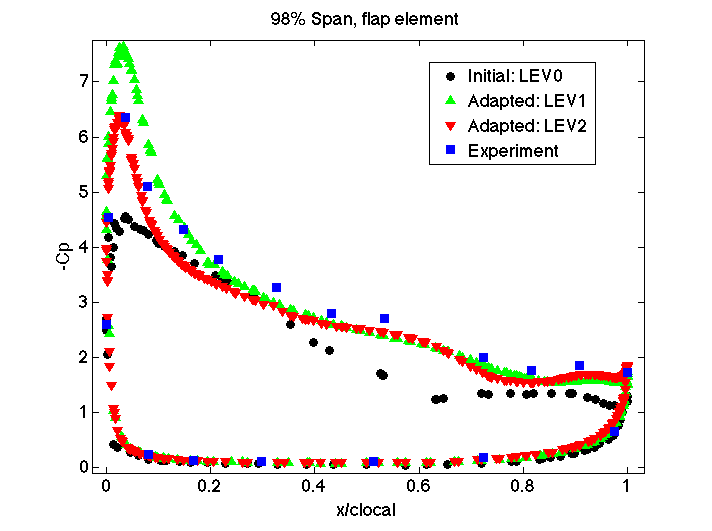}
	\label{f:TWFlap98}
}
\vspace{-10pt}
 \caption{Coefficient of pressure at 17\%, 50\% and 98\% spanwise sections of NASA trap wing}
 \label{f:TWCp1750All}
 \end{center}
\end{figure}

The coefficient of pressure for the slat, the main wing and the flap elements are plotted in Figure~\ref{f:TWCp1750All}. Note that the y-axis is for $-C_p$. For the slat element, the agreement with experiments is good for all the meshes especially for the 17\% and 50\% sections, however, the adapted meshes show less scatter in the $C_p$ values at the 17\% section. Near the tip at 98\% section, the adapted meshes show better agreement with the experiments compared to the initial mesh. 

For the main wing, the agreement is good for all the three meshes with the experimental data at 17\% and 50\% sections. Some small differences are seen near the trailing edge. The initial mesh does not have enough resolution on the pressure side near the trailing edge and thus fails to capture the effect of reduction in the thickness of the wing in that region, which is captured by the adapted meshes. Again, near the tip (Figure~\ref{f:TWMain98}), the initial mesh under predicts separation which leads to a flatter $C_p$ curve for the suction side of the wing, far off from the experimental values, with the worst behavior seen near the trailing edge. With adaptivity, this region receives enough refinement and is captured with a greater accuracy. The $ C_p$ values for the adapted meshes are in good agreement with the experiments near the tip of the main wing for both the adapted meshes. For the pressure side also, the adapted meshes perform better where the thickness of the wing drops down. 

For the flap element, the $ C_p$ values at 17\% and 50\% sections show good agreement with the experimental values. This indicates that the initial mesh is adequate to accurately predict the flow phenomenon away from the tip. This is not surprising since the flow is attached near the in-board and mid-board sections. However, near the tip (Figure~\ref{f:TWFlap98}), for example, at the 98\% span section, the initial mesh under predicts the suction peak. Interestingly, the adapted LEV1 mesh shows an overshoot in the suction pressure peak near the nose of the flap, but the adapted LEV2 mesh shows good agreement with the experiments. This indicates that usually at least a couple of adaptivity passes could be required to arrive at an effective mesh. For the rest of the section, the adapted meshes show better agreement with the experimental data, also capturing the wavy nature near the mid-section area. The initial mesh under predicts the separation near the trailing edge which is again predicted better by the adapted meshes.

\subsubsection{Comparison with nested refinement approach}

To display the effectiveness of adaptive refinement approach, we compare it with nested or uniform refinement. In the $1^{\text{st}}$ high-lift prediction workshop \cite{HLPW1}, participants were required to create a coarse, a medium and a fine mesh and their results were compared. The mesh sizes (mesh edge lengths) in each subsequent grids were 1.5 times smaller than the previous one. We use a similar approach but instead we perform a nested refinement operation to reach the next level of the mesh. We then compare results from these meshes with our adapted mesh results. 

In nested refinement (also called uniform refinement), each mesh edge is split in two. This means that each tetrahedron element in the mesh gets refined into 8 new tetrahedra elements. The mixed element boundary layers are not refined in the direction of the growth curves, meaning that the normal spacing is kept constant. This results in refinement of each prism element into 3 new prism elements and each pyramid element is refined into 2 new pyramid elements. 

\begin{figure}[h!]
\begin{center}
\vspace{10pt}
\subfigure[Initial Mesh: LEV0]{
	\includegraphics[width=7cm]{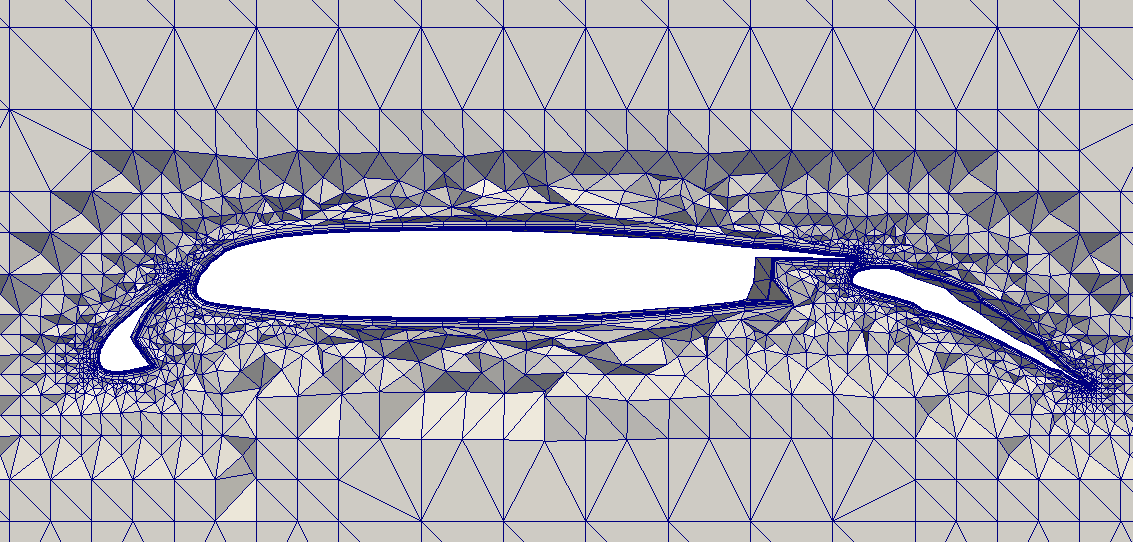}
	\label{f:TWInitMesh}
}
\subfigure[Nested refinement mesh: NLEV1]{
	\includegraphics[width=7cm]{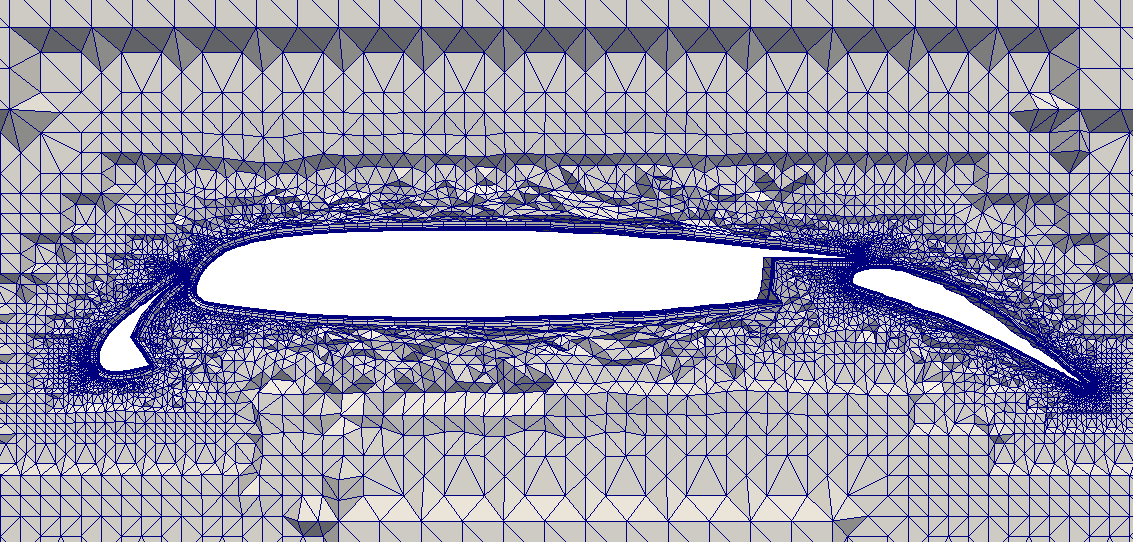}
	\label{f:TWRefine1Mesh}
}	
\subfigure[Nested refinement mesh: NLEV2]{
	\includegraphics[width=7cm]{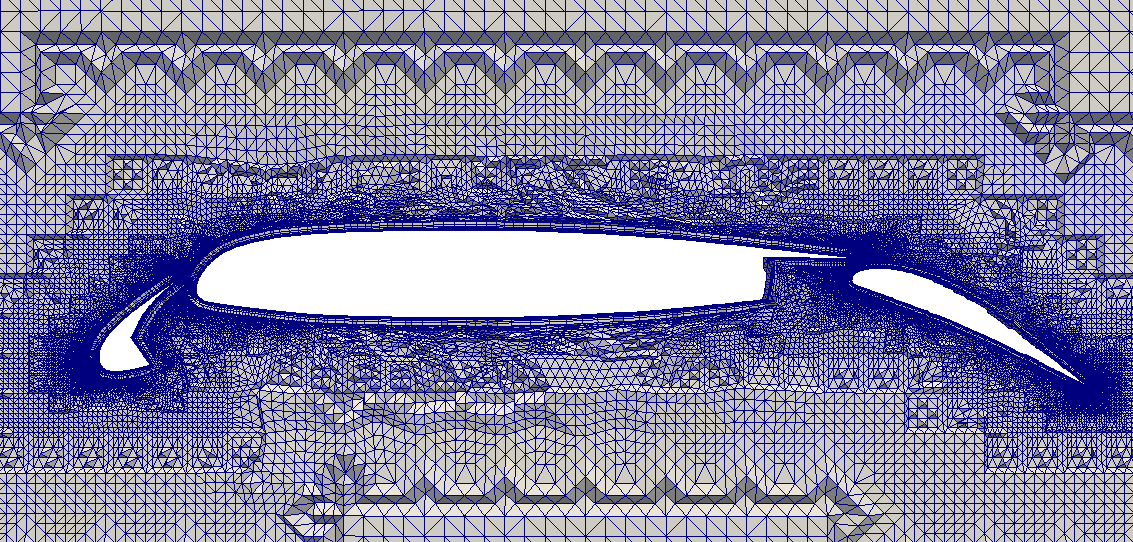}
	\label{f:TWRefine2Mesh}
}
\subfigure[Adapted mesh: LEV2]{
	\includegraphics[width=7cm]{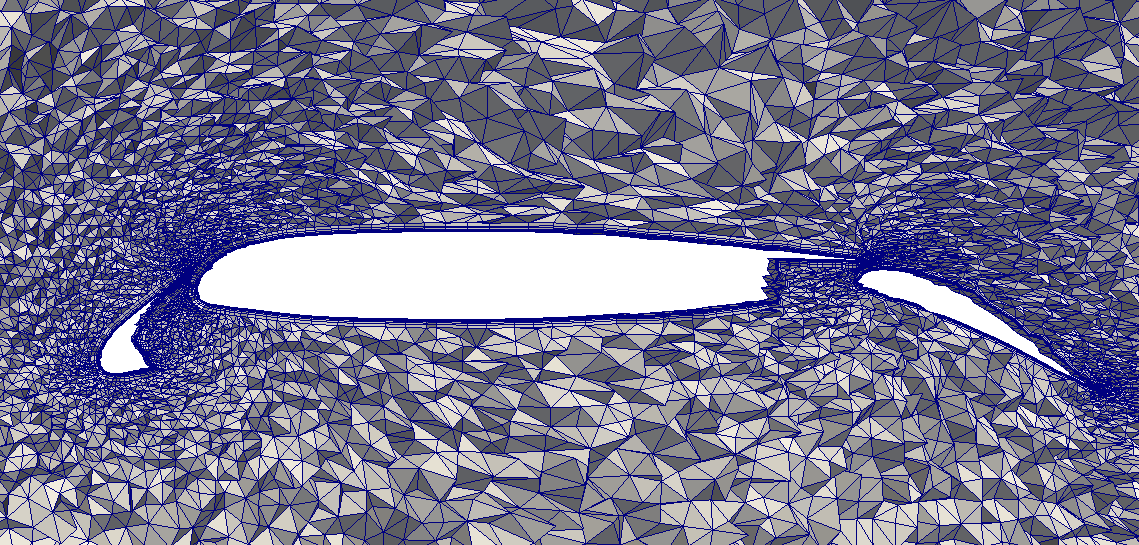}
	\label{f:TWAdapt2Mesh}
}
\vspace{-10pt}
 \caption{Cut views of the initial and the nested refinement meshes for NASA trap wing}
 \label{f:TWURMesh}
\end{center}
\end{figure}

Figure~\ref{f:TWURMesh} shows the cut views of the meshes at each refinement level created with nested refinement and the final (second) adapted mesh for comparison. The uniform refinement of prisms and tetrahedra as explained before can be easily seen in the pictures. Clearly, the adapted mesh shows a lot of anisotropy in the mesh elements which is a major reason for its efficiency over the uniform refinement approach. The refinement is also fairly selective as seen near the suction of the wing and trailing edges of all the components. 
 
\begin{table}[h]
\vspace{10pt}
\caption[Computational comparison of meshes for NASA trap wing]{Summary of initial, final (second) adapted and nested refinement meshes for NASA trap wing}
\newcolumntype{A}{>{\centering\arraybackslash}m{5 cm}}
\newcolumntype{B}{>{\centering\arraybackslash}m{3 cm}}
\begin{tabular}{|A|B|B|}
\hline
 Mesh & \# elements & \# vertices \\
\hline
Initial mesh: LEV0 & 3.39M & 1.13M  \\
\hline
Nested refinement mesh: NLEV1 & 20.65M & 6.65M\\
\hline
Nested refinement mesh: NLEV2 & 139.33M & 37.18M\\
\hline
Adapted mesh: LEV2 & 12.85M & 3.46M \\
\hline
\end{tabular}
\label{t:TrapwingMeshes}
\end{table}

Table~\ref{t:TrapwingMeshes} shows the comparison of the meshes obtained with nested refinement with the initial and the adapted mesh. Between the LEV0 to the refined NLEV2 meshes, the number of elements roughly increases by a factor of 6-7 at each level, whereas the adapted LEV2 mesh has increased only by a factor of 3 after two mesh adaptation cycles. This indicates significant savings in the computational resources if the results of these meshes are similar, which we will investigate next. 

\begin{figure}[h!]
\begin{center}
\subfigure[Slat element: 98\% span]{
	\includegraphics[width=5.7cm]{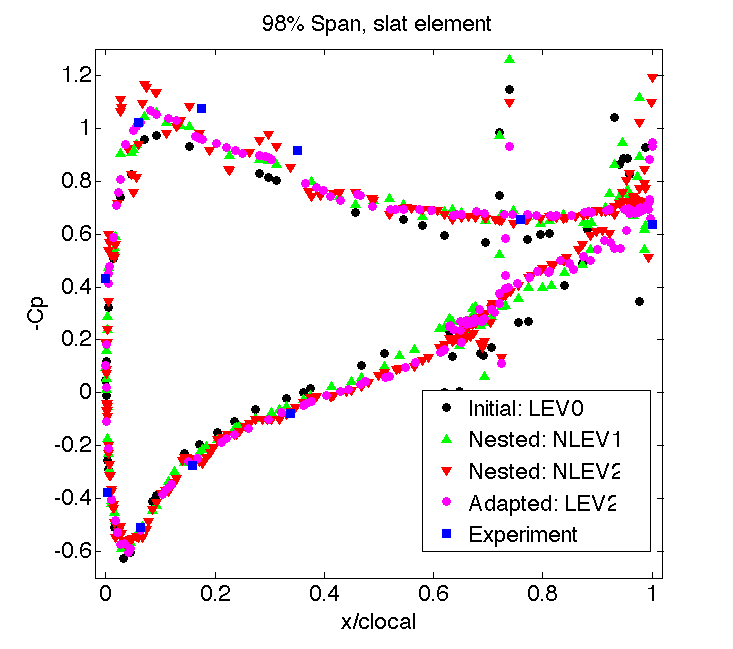}
	\label{f:TWURSlat98}
}
\hspace{-25pt}
\subfigure[Main element: 98\% span]{
	\includegraphics[width=5.7cm]{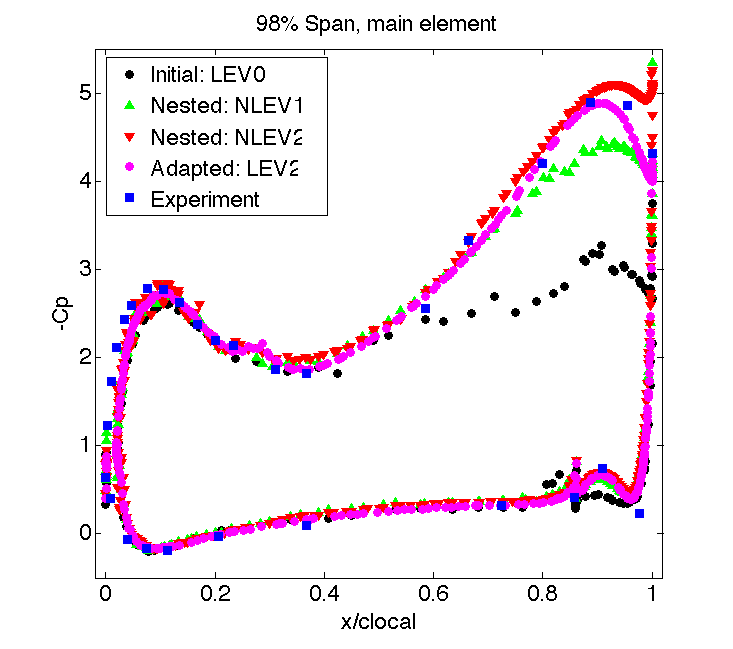}
	\label{f:TWURMain98}
}	
\hspace{-25pt}
\subfigure[Flap element: 98\% span]{
	\includegraphics[width=5.7cm]{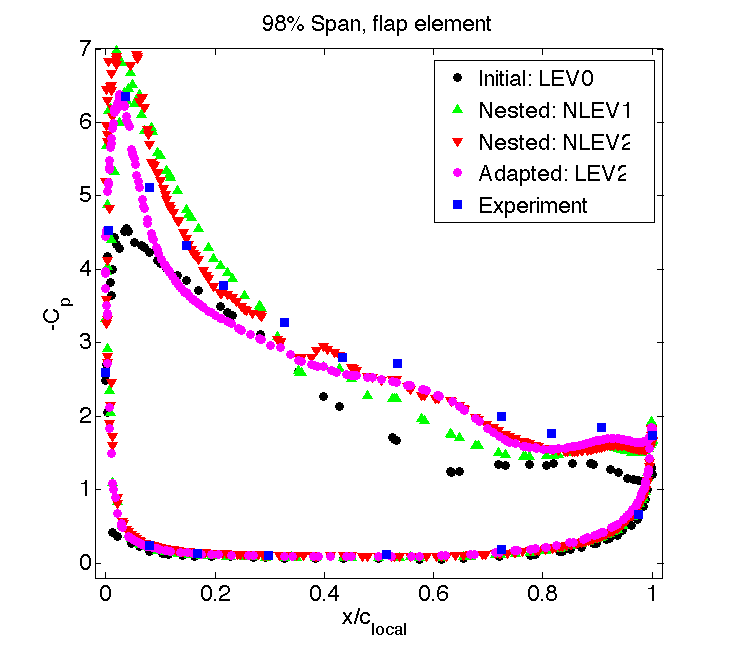}
	\label{f:TWURFlap98}
}
\end{center}
\vspace{-15pt}
 \caption{Coefficient of pressure at 98\% spanwise section for NASA trap wing}
 \label{f:TWURCp}
\end{figure}

Since the $ C_p$ values predicted by the initial mesh agree well with the experiments at 17\% and 50\% spanwise sections, we only plot the $ C_p$ values for the 98\% span section in Figure~\ref{f:TWURCp}, for the meshes created with nested refinement. As the mesh is uniformly refined, the $ C_p$ values gradually start showing better agreement with the experiments, with NLEV2 mesh showing the best results. The adapted LEV2 and the refined NLEV2 mesh both give comparable $ C_p$ values and are in reasonable agreement with the experiments for all the three wing elements. 

\begin{figure}[h!]
\begin{center}
\subfigure{
	\includegraphics[width=7cm]{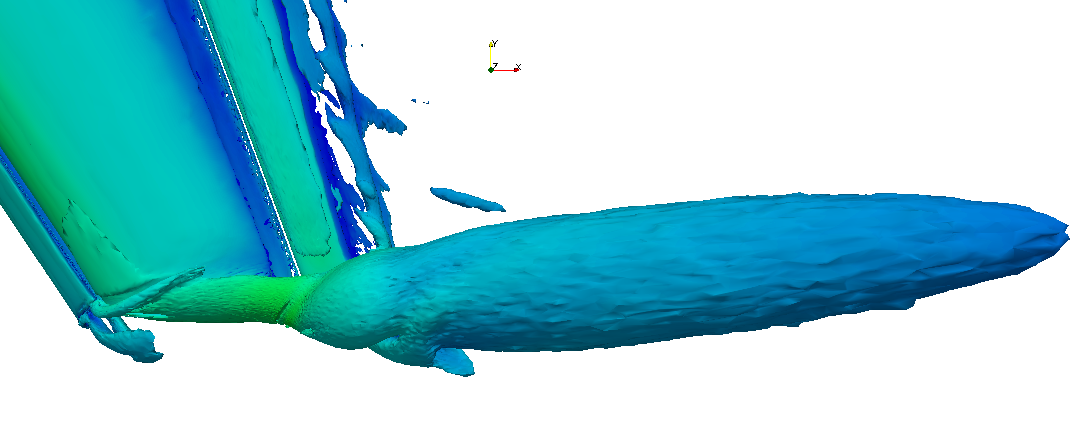}
	\label{f:TWQIsoInit}
}
\subfigure{
	\includegraphics[width=7cm]{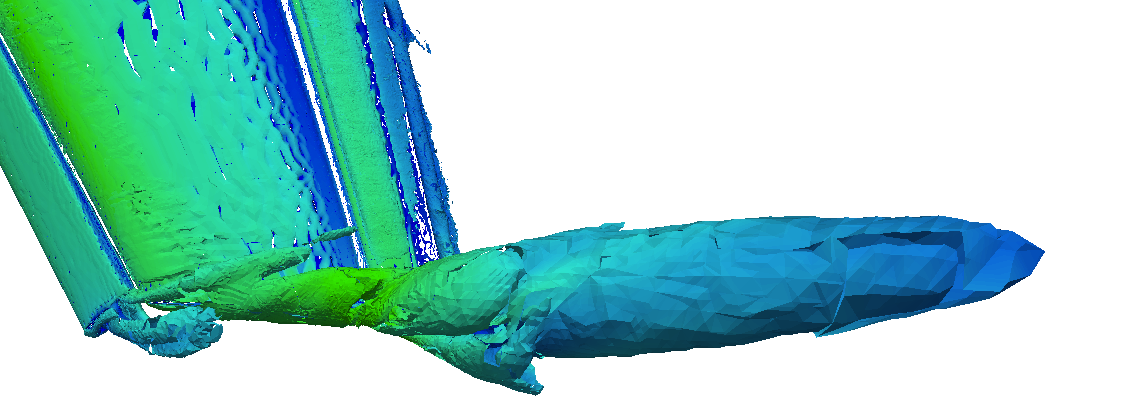}
	\label{f:TWQIsoAdapt2}
}
\subfigure[Adapted mesh: LEV2] {
	\includegraphics[width=5cm]{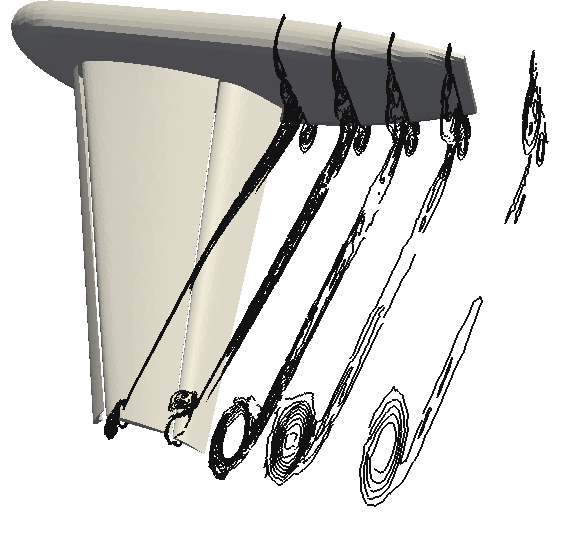}
	\label{f:TWVortContRefine2}
}
\hspace{45pt}
\subfigure[Nested refinement mesh: LEV2]{
	\includegraphics[width=5cm]{TrapWing_TipVortContours_Adapt2.png}
	\label{f:TWVortContAdapt2}
}
\vspace{-10pt}
 \caption{On top: Iso-surfaces of Q criterion colored by speed for NASA trap wing showing the tip vortex, on bottom: contours of vorticity on various slices with normals in the flow direction for NASA trap wing}
 \label{f:TWQIsoUR}
\end{center}
\end{figure}

Figure~\ref{f:TWQIsoUR} shows iso-surfaces for the Q criterion for the NLEV2 mesh and the adapted LEV2 meshes. The shape of the tip vortex is similar for both of the meshes. The size of the vortex given by the NLEV2 mesh is smaller than that of the LEV2 mesh, indicating that the adaptation gives marginally better resolution in this area over the uniformly refined mesh and one more uniform refinement cycle might be required to arrive at the same result as the LEV2 mesh. From the vorticity contours, it can be seen that both of the LEV2 and NLEV2 meshes give similar resolution in the tip area. The off-body vortices are captured to a satisfactory degree by both of these meshes. 

\subsection{EUROLIFT DLR-F11 high lift configuration} 

The second test case for in-plane adaptivity of multi-element wings is the DLR-F11 high lift configuration, which was the focus of the $2^{\text{nd}}$ high-lift prediction workshop\cite{HLPW2}. The configuration is a multi-element wing similar to NASA trap wing, but more complex and with a bigger fuselage. Table~\ref{t:DLRSetup} provides details of the case setup. Settings for adaptation are similar to NASA trap wing. The initial mesh has 10.1 million elements and the adapted mesh contains about 40.69 million elements. The experimental results are explained in Rudnick et al.~\cite{DLRExp}.

\begin{table}[h!]
\centering
\newcolumntype{A}{>{\centering\arraybackslash}m{3 cm}}
\newcolumntype{B}{>{\centering\arraybackslash}m{4 cm}}
\newcolumntype{D}{>{\centering\arraybackslash}m{2 cm}}
      \begin{tabular}{|A|B|A|A|}
      \hline
  	 Mach number & Mean aerodynamic chord (MAC) & $\mathrm{Re_{MAC}}$ & Angle of attack  \\ \hline
	0.175 &  0.347 $ m$ & 15.1 million & 7\degrees  \\ \hline
      \end{tabular}
  \caption{Problem definition for DLR-F11 configuration}
  \label{t:DLRSetup}
\end{table}

\begin{figure}[h!]
\begin{center}
\subfigure{
	\includegraphics[width=7.5cm]{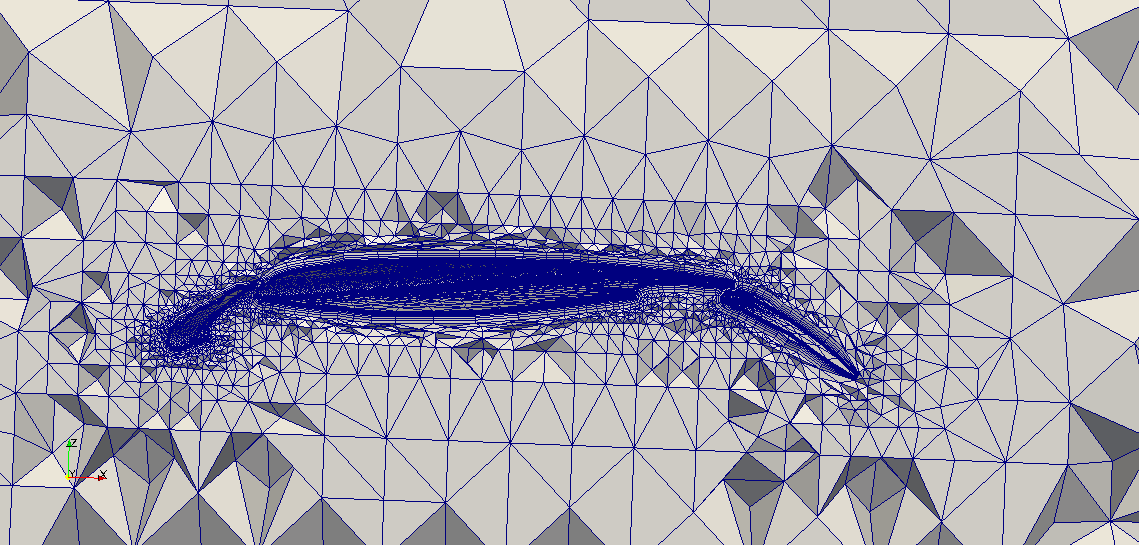}
	\label{f:DLRConfig2}
}
\subfigure{
	\includegraphics[width=7.5cm]{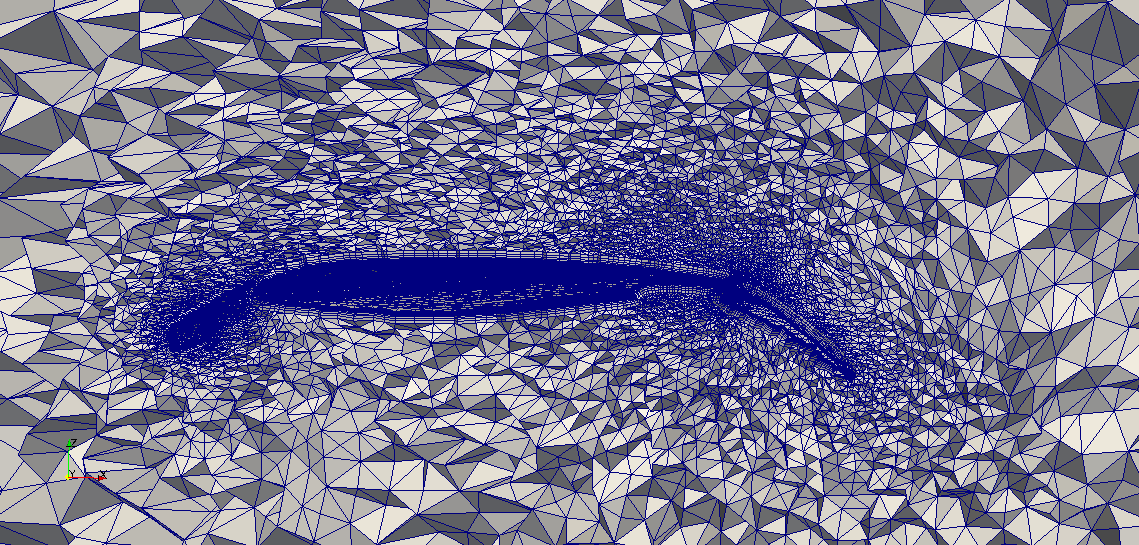}
	\label{f:DLRConfig4}
}
 \subfigure[Initial: LEV0]{
	\includegraphics[width=7.5cm]{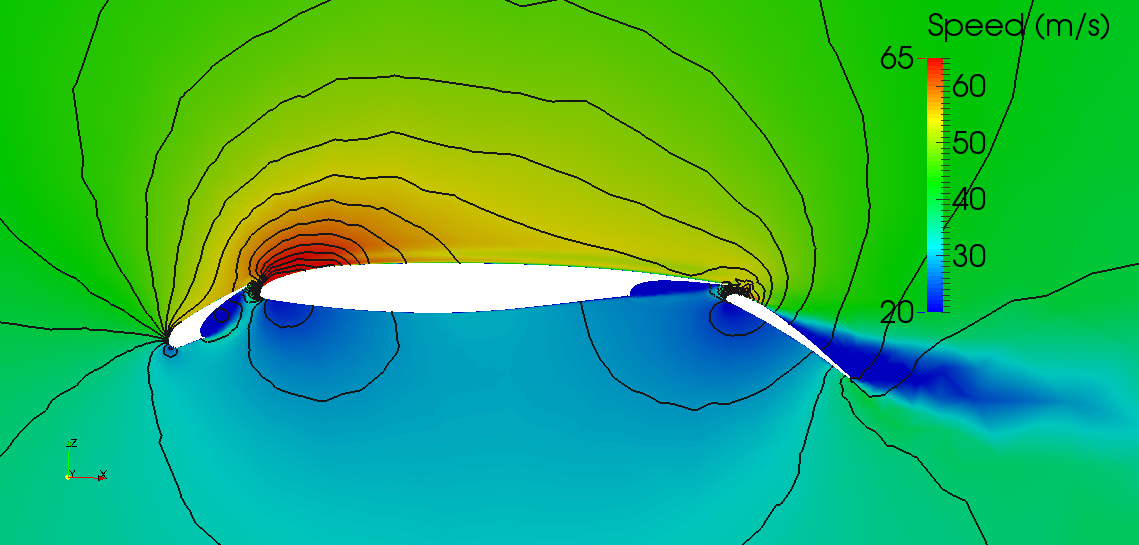}
	\label{f:DLRConfig2}
}
\subfigure[Adapted: LEV1]{
	\includegraphics[width=7.5cm]{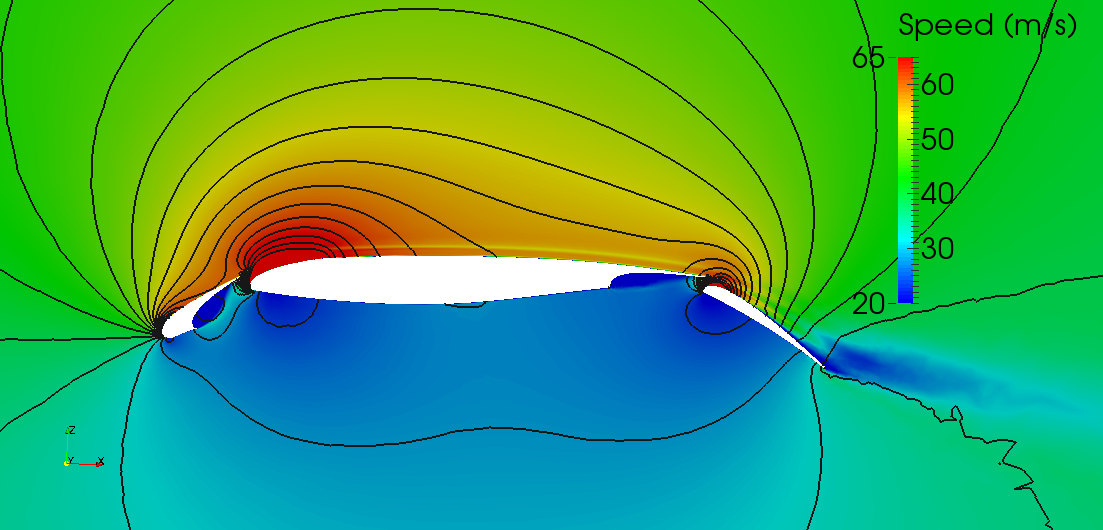}
	\label{f:DLRConfig4}
}
\vspace{-10pt}
 \caption{Cut views of meshes and speed contours (colored) and pressure contours (black lines) for different meshes of DLR-F11}
 \label{f:DLRSpeed}
\end{center}
\end{figure}

Figure~\ref{f:DLRSpeed} shows cut view pictures of the meshes at mid-span section and their corresponding speed contours (in color) and pressure contours (black lines). The pressure contours are jagged and irregular for the initial coarse mesh, but become smoother for the adapted mesh due to an increase in the mesh resolution. As a general trend the initial mesh shows more separation than the adapted mesh over the flap element. This effect is also captured by the coefficient of pressure plots on the flap sections. 

\begin{figure}[h!]
\begin{center}
\subfigure[Slat element: 29\% span]{
	\includegraphics[width=5.7cm]{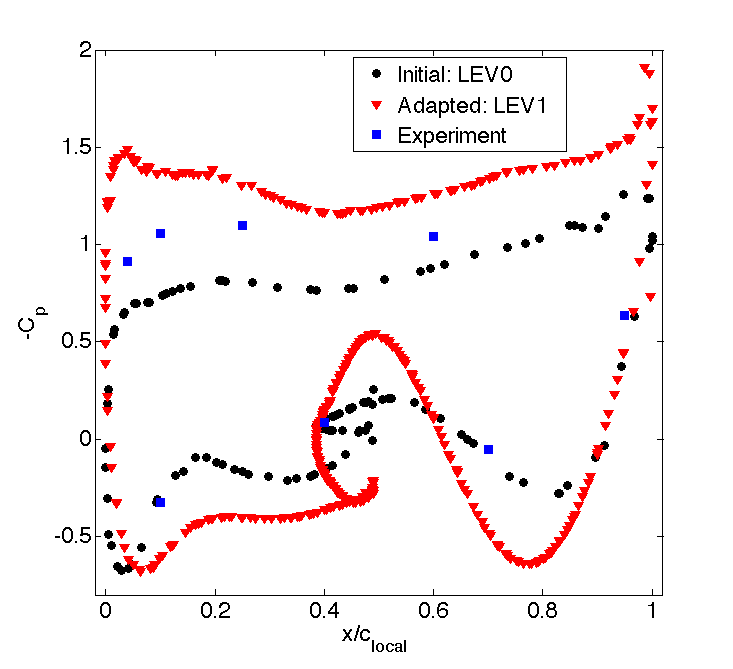}
	\label{f:DLRSlat98}
}
\hspace{-25pt}
\subfigure[Slat element: 68\% span]{
	\includegraphics[width=5.7cm]{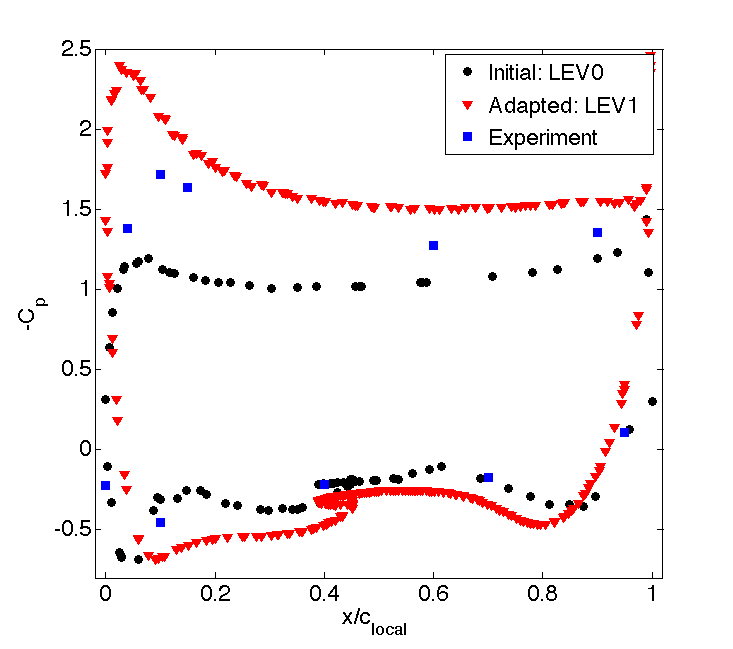}
	\label{f:DLRMain98}
}	
\hspace{-25pt}
\subfigure[Slat element: 90\% span]{
	\includegraphics[width=5.7cm]{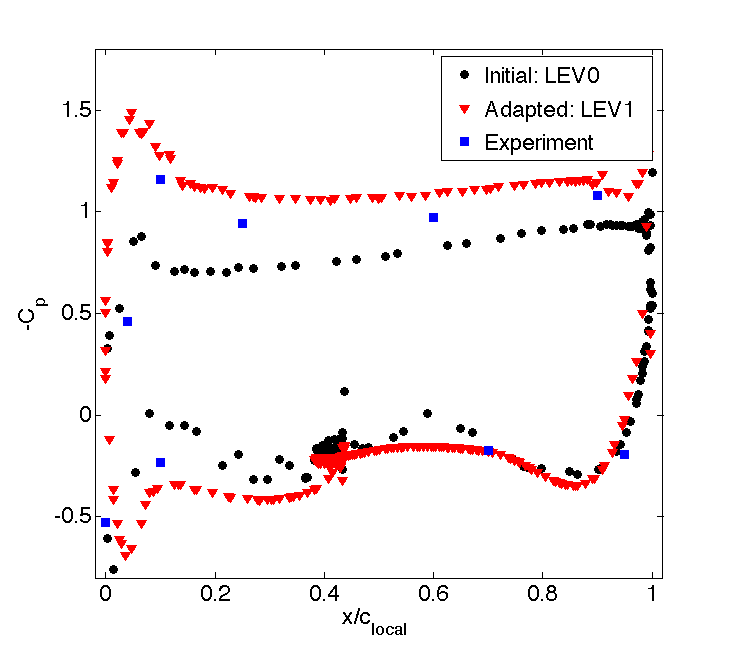}
	\label{f:DLRFlap98}
}
\vspace{-5pt}

\subfigure[Main element: 29\% span]{
	\includegraphics[width=5.7cm]{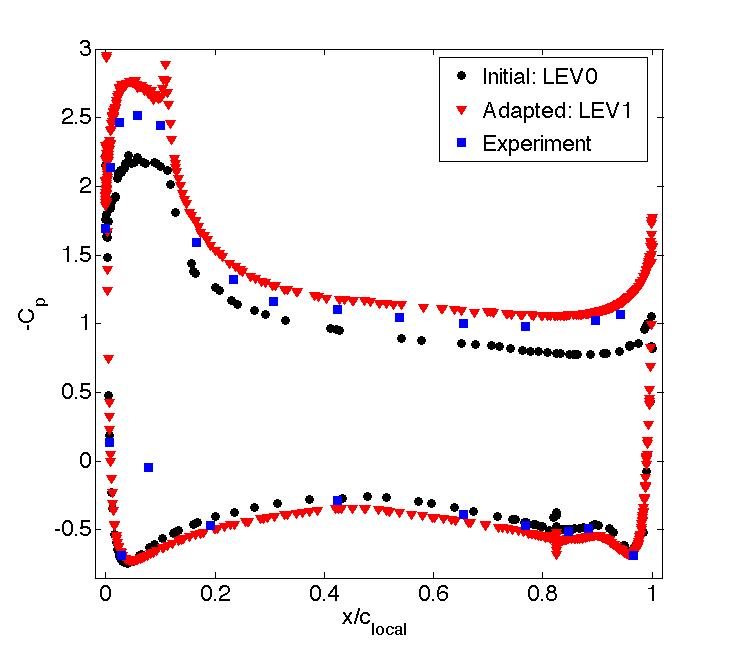}
	\label{f:DLRSlat98}
}
\hspace{-25pt}
\subfigure[Main element: 68\% span]{
	\includegraphics[width=5.7cm]{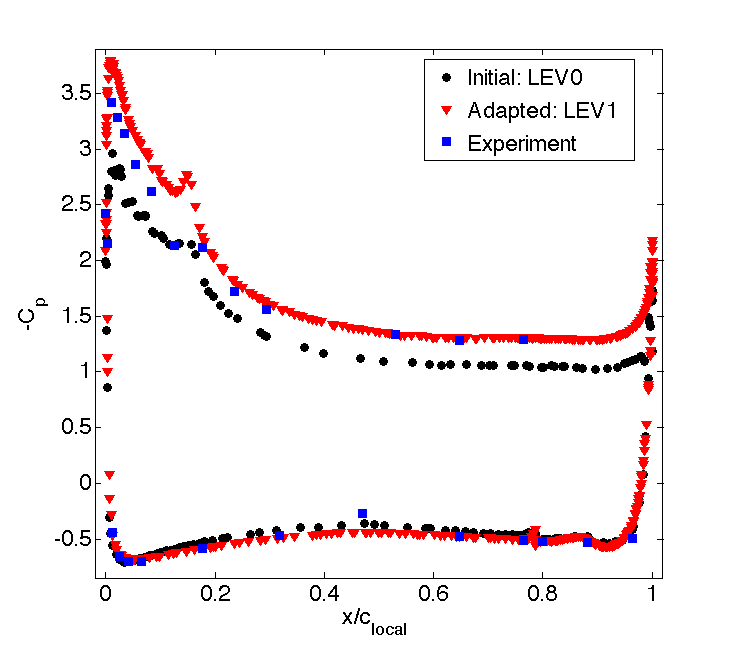}
	\label{f:DLRMain98}
}	
\hspace{-25pt}
\subfigure[Main element: 90\% span]{
	\includegraphics[width=5.7cm]{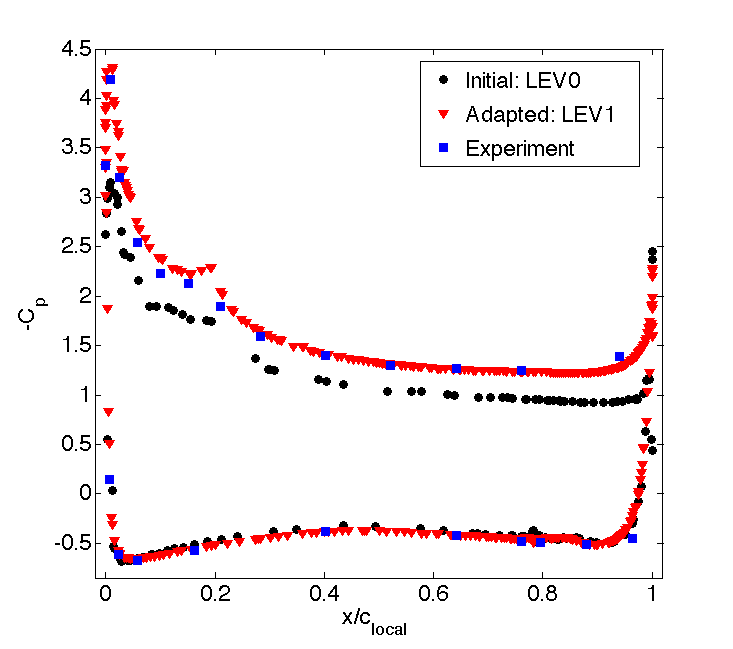}
	\label{f:DLRFlap98}
}
\vspace{-5pt}

\subfigure[Flap element: 29\% span]{
	\includegraphics[width=5.7cm]{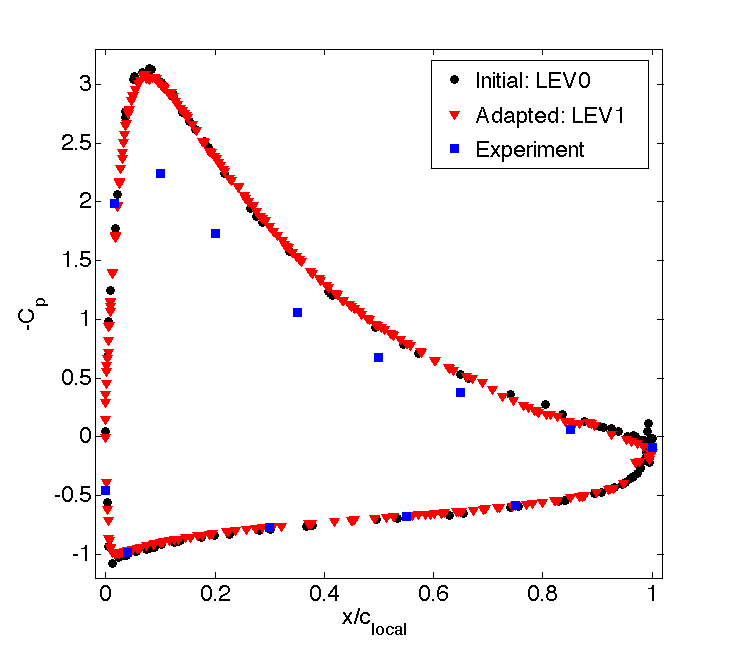}
	\label{f:DLRSlat98}
}
\hspace{-25pt}
\subfigure[Flap element: 68\% span]{
	\includegraphics[width=5.7cm]{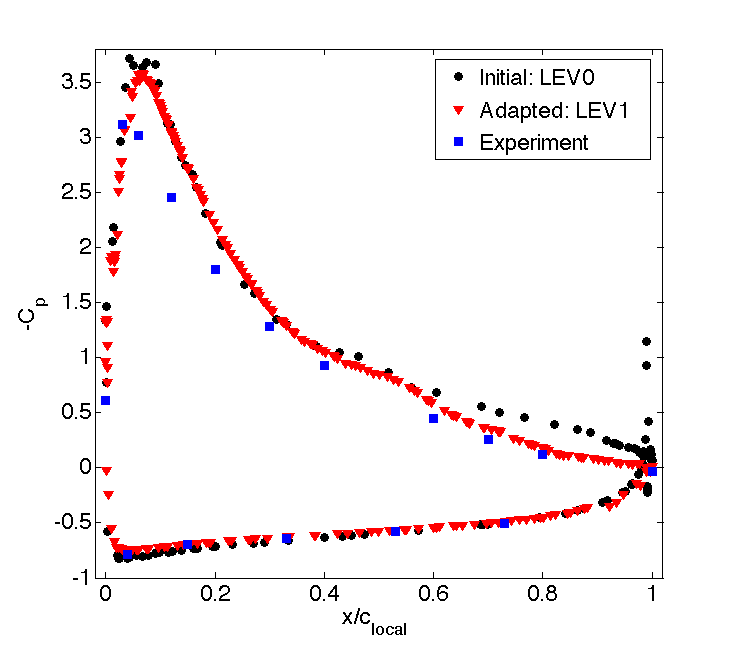}
	\label{f:DLRMain98}
}	
\hspace{-25pt}
\subfigure[Flap element: 90\% span]{
	\includegraphics[width=5.7cm]{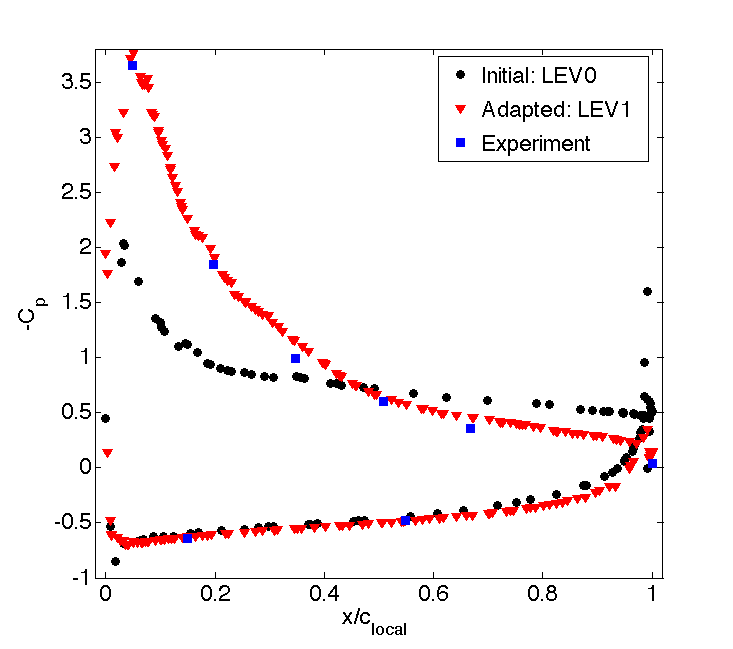}
	\label{f:DLRFlap98}
}
\vspace{-10pt}
 \caption{Coefficient of pressure at 29\%, 68\% and 90\% spanwise sections of DLR-F11}
 \label{f:DLRCpAll}
\end{center}
\end{figure}

Figure~\ref{f:DLRCpAll} plots coefficient of pressure for DLR-F11 case at different sections of the wing. Overall the adapted mesh is able to predict $ C_p$ in better agreement with experiments than the initial mesh. This difference is most noticeable at 90\% section near the wing tip. For this section, the initial mesh predicts attached flow behavior near the trailing edge of the main wing whereas the adapted mesh predicts flow separation which is in accordance with the experimental data. For the flap element, the initial mesh underpredicts the suction peak whereas the adapted mesh is able to capture the behavior correctly. As a part of the $ 2^{\text{nd}}$ high-lift prediction workshop, coarse, medium and fine meshes were constructed for this geometry. Comparison of adapted mesh results with these progressively refined meshes is given in more details in Chitale et al.\cite{ChitaleHLPW2}.

Table~\ref{t:DLRF11CL} shows comparison of $ C_L$ and $ C_D$ for the initial and the adapted meshes with experimental values. Clearly, the adapted mesh results are closer to the experiments than the initial mesh, especially $ C_L$. Another adaptivity pass might be needed to achieve better agreement with experimental $ C_L$ and $ C_D$ and to show grid convergence. 

\begin{table}[h!]
\centering
\newcolumntype{A}{>{\centering\arraybackslash}m{3 cm}}
\newcolumntype{B}{>{\centering\arraybackslash}m{4 cm}}
\newcolumntype{D}{>{\centering\arraybackslash}m{2 cm}}
      \begin{tabular}{|A|A|A|}
      \hline
  	  & $ C_L$ & $ C_D$   \\ \hline
	Initial: LEV0 &  1.787 & 0.160  \\ \hline
	Adapted: LEV1 &  1.966 & 0.163  \\ \hline
	Experiments & 1.930 & 0.1619\\ \hline

      \end{tabular}
  \caption{Comparison of $ C_L$ and $ C_D$ for DLR-F11}
  \label{t:DLRF11CL}
\end{table}

For the DLR-F11 configuration, one adapt cycle has shown significant improvements over the initial mesh in terms of $ C_L$, $ C_D$ and surface $ C_p$. Another adaptive pass would confirm and verify our results, however, the adapted mesh has nearly 40 million elements and further adaptation would require significantly more computations with a parallel approach, which is beyond the scope of this paper and is left for future studies. 


\section{Conclusion}

In this work we applied and analyzed anisotropic adaptivity on two complex multi-element wing configurations; NASA trap wing and EUROLIFT DLR-F11 configuration. The focus was on in-plane adaptation of the boundary layer meshes. Adaptivity proved to be an efficient way to reach finer grid spacings in important regions like flow separation locations, wing tip area and in wakes. This improved resolution led to more accurate prediction of the flow structures and better agreement with the experimental results. Error indicator driven anisotropic adaptivity also saved important computational resources when compared to nested or uniformly refinement meshes.

\section*{Acknowledgments}

The work was carried out at University of Colorado Boulder, in collaboration with Simmetrix, Inc. and Rensselaer Polytechnic Institute. Support was provided by the NASA STTR Phase II Grant No. BEE103/NNX11CC69C.R. We also gratefully acknowledge Simmetrix Inc. for their meshing and geometric modeling libraries, Kitware for their visualization tools and Altair Engineering (previously ACUSIM, Inc.) for their linear algebra solver library. This work utilized: (i) the Mira BlueGene/Q supercomputer at ANL which is supported by the Department of Energy, (ii) the Janus supercomputer, which is supported by the National Science Foundation (award number CNS-0821794) and the University of Colorado Boulder, and (iii) the computational resources provided by the Center for Computational Innovations (CCI) at Rensselaer Polytechnic Institute.

\newpage

\bibliography{bibtex_database}
\bibliographystyle{aiaa}

\end{document}